\documentclass[nofootinbib,aps,prl,reprint,groupedaddress,superscriptaddress,showpacs]{revtex4-1}
\usepackage[english]{babel}
\usepackage[utf8x]{inputenc}
\usepackage{amssymb}
\usepackage{amsthm}
\usepackage{amsmath}
\usepackage{hyperref}
\usepackage{subfigure}
\usepackage[percent]{overpic} 
\usepackage{picture}
\usepackage{color}
\usepackage{nth}
\usepackage{gensymb}
\usepackage{cleveref}
\usepackage{textcomp}
\usepackage{makecell}
\usepackage{lineno} 
\usepackage{textcomp}

\usepackage{lineno}

\graphicspath{{img/}}

\begin{document}

\title{Longitudinal phase-space manipulation with beam-driven plasma wakefields}


\author{V. Shpakov}
\email[]{vladimir.shpakov@lnf.infn.it}
\author{M.P. Anania}
\author{M. Bellaveglia}
\author{A. Biagioni}
\author{F. Bisesto}
\author{F. Cardelli}
\author{M. Cesarini}
\author{E. Chiadroni}
\affiliation{Laboratori Nazionali di Frascati, Via Enrico Fermi 40, 00044 Frascati, Italy}
\author{A. Cianchi}
\affiliation{University or Rome Tor Vergata and INFN, Via Ricerca Scientifica 1, 00133 Rome, Italy}
\author{G. Costa}
\author{M. Croia}
\author{A. Del Dotto}
\author{D. Di Giovenale}
\affiliation{Laboratori Nazionali di Frascati, Via Enrico Fermi 40, 00044 Frascati, Italy}
\author{M. Diomede}
\affiliation{Sapienza University, Piazzale Aldo Moro 5, 00185 Rome, Italy}
\author{M. Ferrario}
\author{F. Filippi}
\author{A. Giribono}
\author{V. Lollo}
\affiliation{Laboratori Nazionali di Frascati, Via Enrico Fermi 40, 00044 Frascati, Italy}
\author{M. Marongiu}
\affiliation{Sapienza University, Piazzale Aldo Moro 5, 00185 Rome, Italy}
\author{V. Martinelli}
\affiliation{Laboratori Nazionali di Frascati, Via Enrico Fermi 40, 00044 Frascati, Italy}
\author{A. Mostacci}
\affiliation{Sapienza University, Piazzale Aldo Moro 5, 00185 Rome, Italy}
\author{L. Piersanti}
\author{G. Di Pirro}
\author{R. Pompili}
\author{S. Romeo}
\author{J. Scifo}
\author{C. Vaccarezza}
\author{F. Villa}
\affiliation{Laboratori Nazionali di Frascati, Via Enrico Fermi 40, 00044 Frascati, Italy}
\author{A. Zigler}
\affiliation{Laboratori Nazionali di Frascati, Via Enrico Fermi 40, 00044 Frascati, Italy}
\affiliation{Racah Institute of Physics, Hebrew University, 91904 Jerusalem, Israel}

\date{\today}

\begin{abstract}
The development of compact accelerator facilities providing high-brightness beams is one of the most challenging tasks in field of next-generation compact and cost affordable particle accelerators, to be used in many fields for industrial, medical and research applications. The ability to shape the beam longitudinal phase-space, in particular, plays a key role to achieve high-peak brightness. Here we present a new approach that allows to tune the longitudinal phase-space of a high-brightness beam by means of a plasma wakefields. The electron beam passing through the plasma drives large wakefields that are used to manipulate the time-energy correlation of particles along the beam itself. We experimentally demonstrate that such solution is highly tunable by simply adjusting the density of the plasma and can be used to imprint or remove any correlation onto the beam. This is a fundamental requirement when dealing with largely time-energy correlated beams coming from future plasma accelerators.

\end{abstract}


\maketitle

High-brightness electron beams are nowadays used for many applications like, for instance, Inverse Compton Scattering~\cite{schoenlein1996femtosecond,bacci2013electron}, the generation of THz~\cite{chiadroni:022703,giorgianni2016strong}, Free Electron Laser (FEL) radiation~\cite{ackermann2007operation,emma2010first,allaria2012highly,petrillo2013observation} and for new plasma-based acceleration techniques~\cite{1979PhRvL..43..267T,1987PhRvL..58..555R,rosenzweig1991acceleration,litos2014high}.
The generation of such beams always require manipulations of their longitudinal phase-space (LPS) in order to achieve peak currents as large as required by the specific task. The ability to shape the energy and temporal profiles is thus of paramount importance.
In FEL facilities, for instance, peak currents of several kA are produced by longitudinally compressing a time-energy correlated (i.e. \textit{chirped}) beam in a dispersive magnetic chicane, where the path length is energy dependent~\cite{giannessi2011self,allaria2012highly}.
The manipulation of the LPS is also a fundamental step in view of the development of new compact machines that exploit advanced acceleration techniques based on plasma wakefields. In this case accelerating fields up to tens of GV/m, $\sim2-3$ orders of magnitude larger than conventional radio-frequency (RF) structures, have been demonstrated allowing to produce GeV level beams in few centimeters~\cite{2007Natur.445..741B,litos2014high,leemans2006gev,faure2006controlled}.
However, due to the shortness of the accelerating field wavelength a large correlated energy spread is imprinted on the accelerated beam, making difficult to transport the beam using conventional magnetic optics (like solenoids and quadrupoles), due to chromatic effects. In this case, a technique able to remove such an energy-chirp must be foreseen.

In this Letter we discuss a new approach that allows to tune the beam LPS by using the wakefields excited in a plasma channel. Other techniques based on the use of metallic~\cite{piot2003longitudinal,england2005sextupole} or dielectric structures~\cite{antipov2014experimental,bettoni2016temporal,penco2017passive} have been also demonstrated. However, in the first case the imprinted energy-chirps cannot exceed few MeV/m while in the second one the tunability is rather limited, depending on the aperture and size of the employed devices.

\begin{figure}[!h]
\centering
\includegraphics[width=0.9\linewidth]{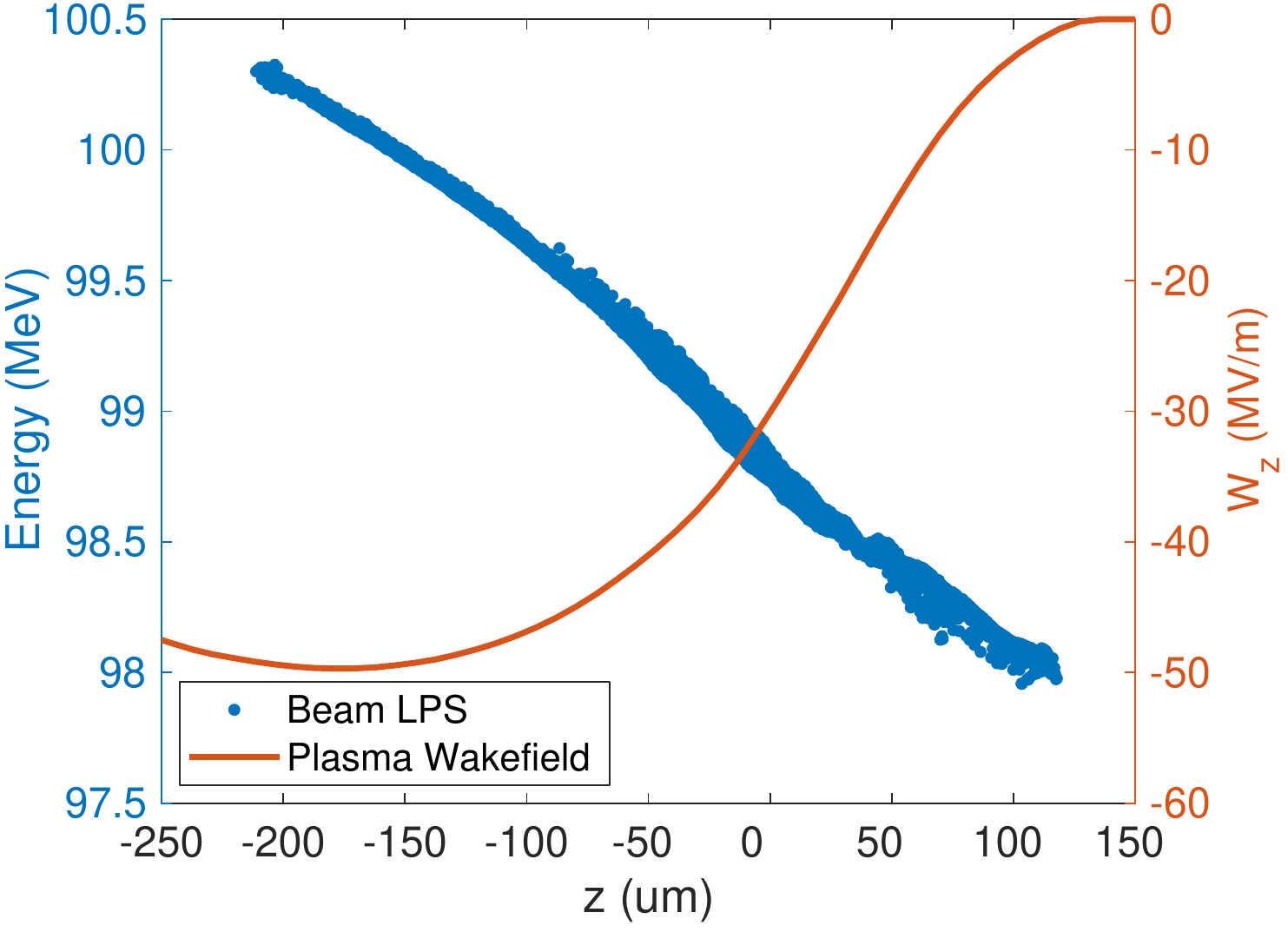}
\caption{LPS of the beam and longitudinal plasma wakefield $W_z$ (red line) produced into a plasma with density $n_p=1.6\times 10^{14}$~cm$^{-3}$ by a moving electron bunch (blue dots).}
\label{GPT_LPS_Wake}
\end{figure}

\begin{figure*}[t]
\centering
\includegraphics[width=0.9\linewidth]{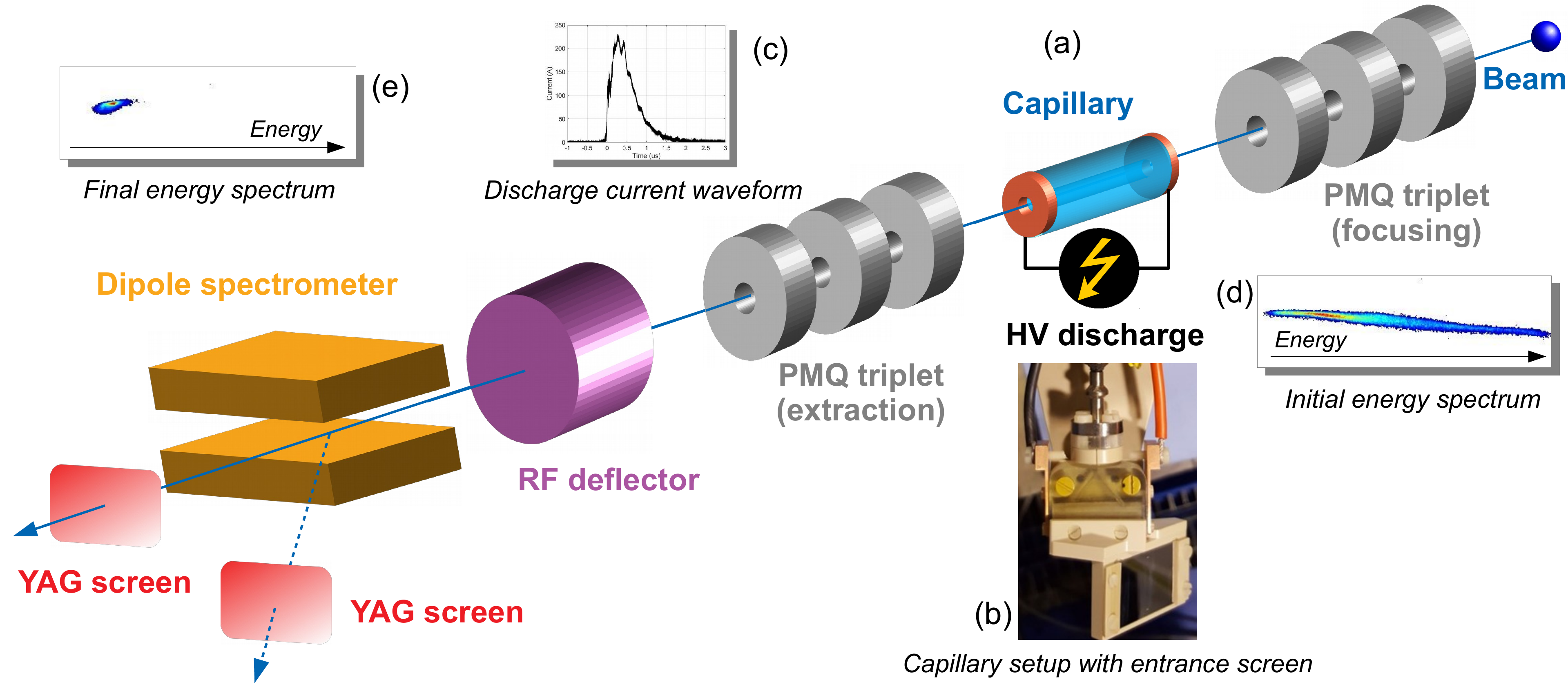}
\caption{Experimental setup. The electron beam is tightly focused by the PMQ triplet into a 3~cm-long plastic capillary (a) filled by H$_2$ gas through two inlets (b) connected to an electrolytic generator. Below the capillary and in correspondence of its entrance, an OTR screen has been installed to measure the beam transverse profile. At the capillary ends there are two copper electrodes connected to a 20~kV power supply producing 230~A peak current (c). The whole system is mounted on a movable actuator allowing to adjust its position with respect to the beam. The exiting beam is then captured by a second PMQ triplet. The diagnostics of the experiment is completed by a RF-deflector and two Ce:YAG screens downstream the magnetic spectrometer. The second screen is located at $14\degree$ with respect to the initial beam path, allowing to measure the beam energy spectrum without (d) and with (e) plasma.}
\label{CapillarySetup}
\end{figure*}

Our solution is based on the use of the self wakefields created by the beam in the plasma and can be employed both to remove the energy-chirp (acting like a \textit{dechirper}) or tune it by adjusting the plasma density \cite{wupreliminary, d2019tunable}.
The basic idea of the LPS manipulation is shown in Fig.~\ref{GPT_LPS_Wake}, where we show the LPS and computed plasma wakefield (red line) produced by a 200~pC bunch in a plasma whose density is $n_p=1.6\times 10^{14}$~cm$^{-3}$. By indicating the energy deviation of each particle along the bunch as $E(z)\approx E_0+h_1 z$, with $h_1$ the first order chirp term, the reported bunch has a negative chirp (higher energy particles on the tail) of $h_1\approx -8\times 10^3$~MeV/m with an overall head-to-tail energy offset of $\approx 2$~MeV ($\Delta E/E\sim 2\%$).
Once injected into the plasma, the electron bunch starts to create the wakefield. Strength of that field depends on plasma density and the density of the beam itself~\cite{fang2014effect}. 
In our configuration the tail of the beam experiences a decelerating electric field and looses its energy while the head moves along an unperturbed plasma, keeping its energy actually constant. It is equivalent to a rotation of the beam LPS and, being induced by a wakefield approximately 50~MV/m, we expect that the energy-chirp can be completely removed by employing a few cm-long plasma structure.

The experiment has been performed at the SPARC\_LAB test-facility~\cite{ferrario2013sparc_lab,pompili2018recentres} by employing 3~cm-long discharge-capillary filled by Hydrogen gas~\cite{PhysRevLett.121.174801,lens_alberto,pompili2017experimental}.
The experimental setup is shown in Fig.~\ref{CapillarySetup}. The bunch is produced by the SPARC photo-injector~\cite{Alesini2003345,chiadroni2013sparc}, consisting of a 1.6~cell RF-gun~\cite{cianchi2008high} followed by two S-band accelerating sections embedded in solenoids coils~\cite{ferrario2010experimental,pompili2016beam}) and one C-band structure.
The plasma device consists of a plastic capillary with length $L_c=3$~cm and $R_c=1$~mm hole radius. The capillary is filled at 1~Hz rate with H$_2$ gas (produced by an electrolytic generator) through two inlets placed at $L_c/4$ and $3L_c/4$ and has two electrodes at its extremities connected to the discharge circuit with a 20~kV pulser~\cite{anania2016plasma} and able to provide 230~A peak discharge-current with shot-to-shot fluctuation $<10$~ns~\cite{PhysRevLett.121.174801}. The peak plasma density reached in the capillary is $n_p\approx 3\times 10^{16}$~cm$^{-3}$, estimated by measuring the H$_{\beta}$ Balmer line with a Stark broadening-based diagnostics~\cite{filippi2016spectroscopic}.
The capillary is installed in a vacuum chamber directly connected to a photo-injector by a windowless, three-stage differential pumping system, that ensures $10^{-8}$~mbar pressure in the RF linac while flowing H$_2$ into the capillary. This solution allows to avoid using any window, thus preventing the beam emittance deterioration by multiple scattering.

\begin{figure}[h]
\centering
\includegraphics[width=0.9\linewidth]{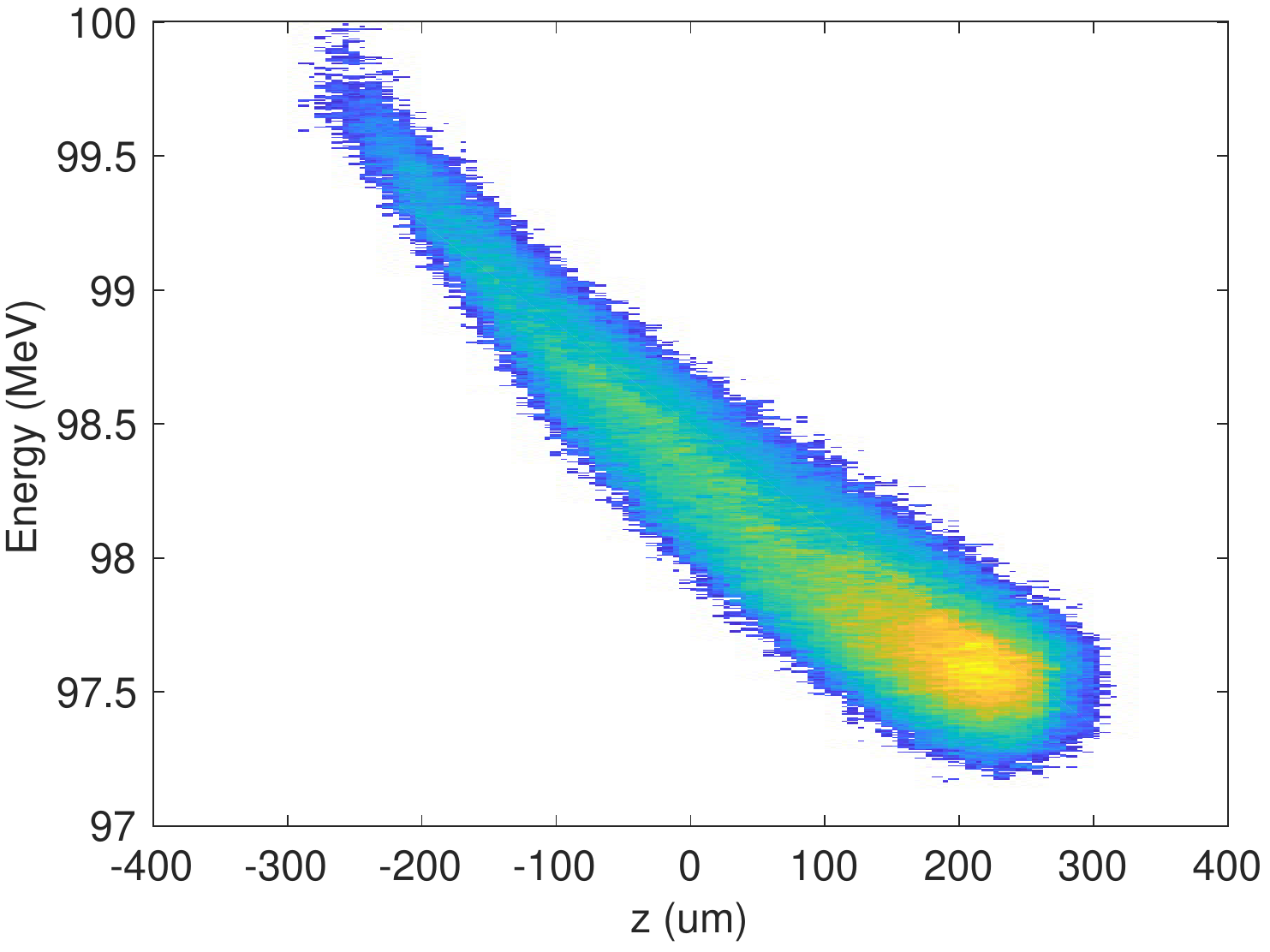}
\caption{Experimentally measured LPS of the negatively chirped bunch. The (rms) energy spread and duration are $\sigma_E\approx 0.6$~MeV and $\sigma_t\approx 250$~fs (corresponding to $\sigma_z\approx 75~\mu m$), respectively. This measurement is obtained without any plasma in the capillary.}
\label{LPSbeam}
\end{figure}

To experimentally produce a chirped LPS, like the one simulated in Fig.~\ref{GPT_LPS_Wake}, we have used the first linac accelerating section as RF compressor by means of the velocity-bunching (VB) technique~\cite{serafini2001velocity,ferrario2010experimental}, that allows to shorten the beam and imprint an energy-chirp on it~\cite{eos_jitter,pompili2016beam}. The induced chirp is negative ($h_1<0$) till the maximum compression point (shortest bunch length) is reached.
Figure~\ref{LPSbeam} shows the measured LPS of the resulting beam. The electron bunch has 200~pC charge, 100~MeV energy (0.6~MeV energy spread) and 250~fs duration (corresponding to $\sigma_z\approx 75~\mu m$ length), measured with a RF-Deflector device~\cite{alesini2006rf}. Its normalized emittance on the horizontal (vertical) plane is $\epsilon_{x(y)}\approx 1.1(1.4)~\mu m$. A triplet of permanent-magnet quadrupoles (PMQ)~\cite{pompili2018compact} allows to squeeze the beam transverse size down to $\sigma_{x(y)}\approx 20(32)~\mu m$. All these quantities are quoted as rms.
An almost linear negative chirp ($h_1\approx -8\times 10^3$~MeV/m) is achieved by moving the RF-phase of the compressor $4\degree$ before the maximum compression point.

\begin{figure}[h]
\centering
\subfigure{
\begin{overpic}[width=0.9\linewidth]{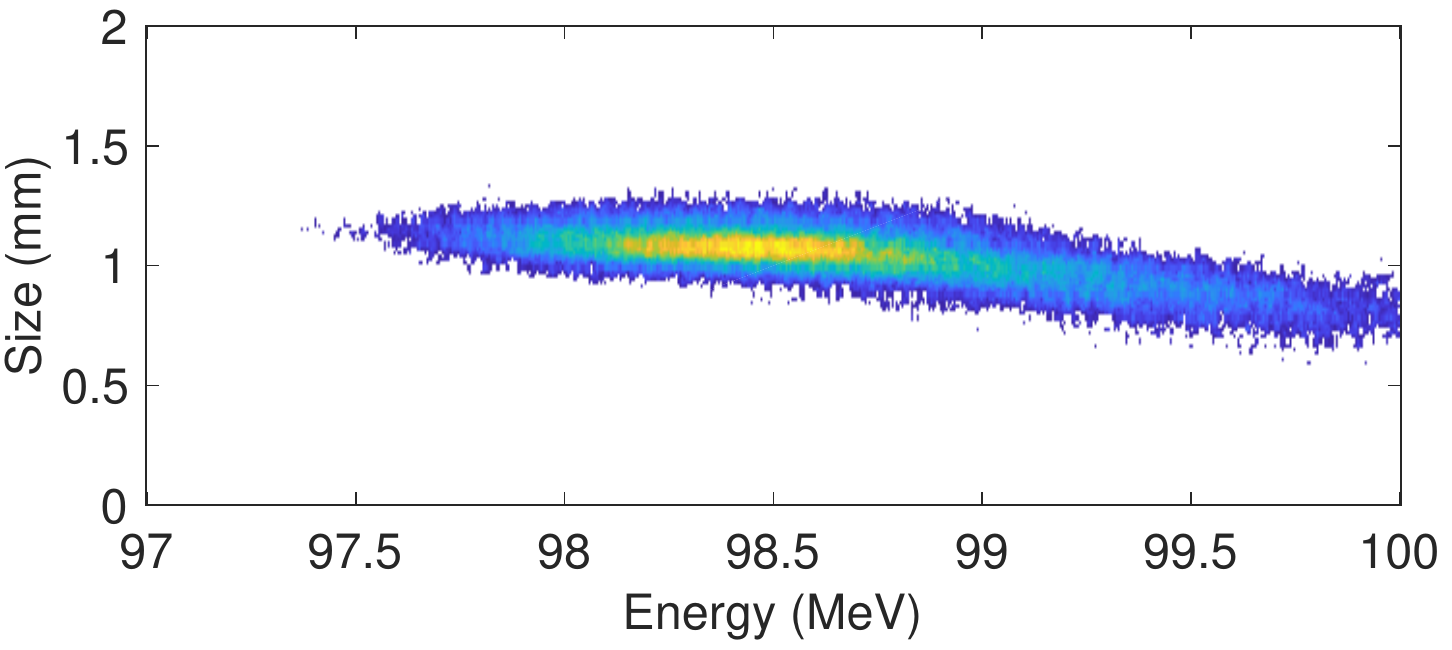}
\put(12,38){\color{black}\textbf{a}}
\end{overpic}
\label{EnergyNoPlasma}
}
\subfigure{
\begin{overpic}[width=0.9\linewidth]{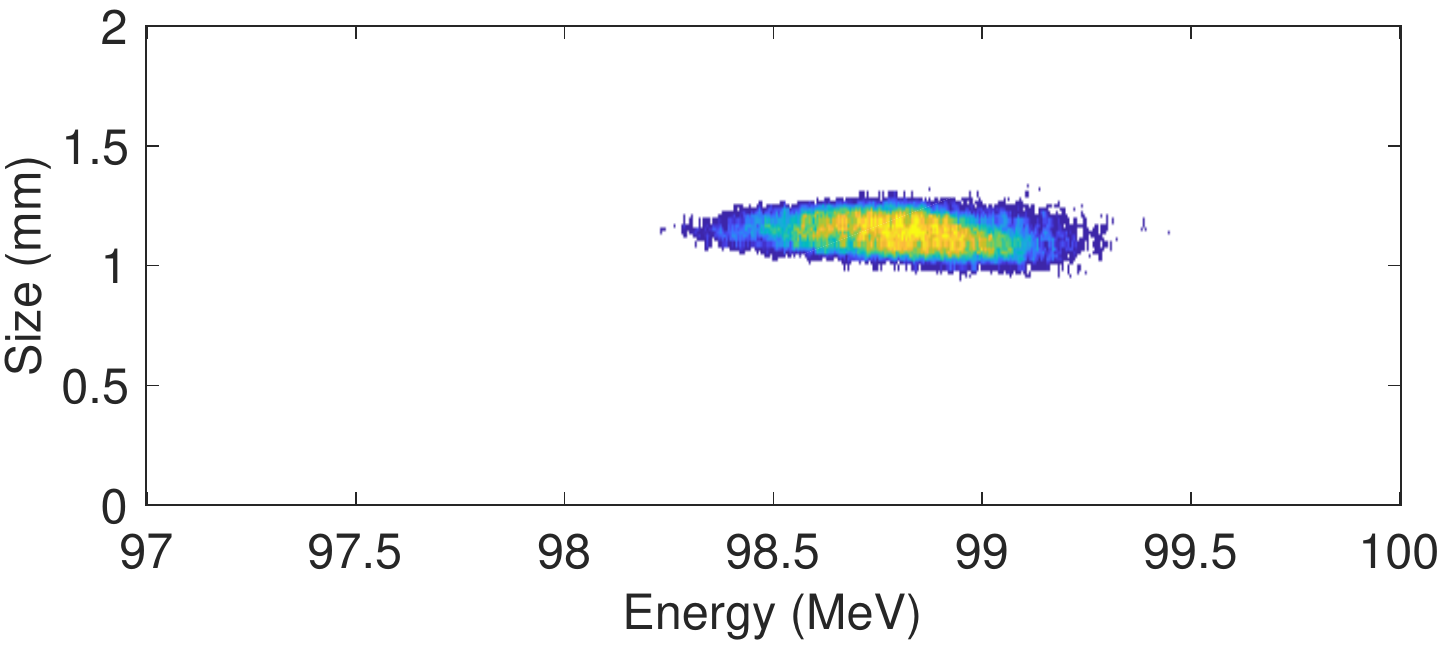}
\put(12,38){\color{black}\textbf{b}}
\end{overpic}
\label{EnergyWithPlasma}
}
\caption{Energy spectrum of the negatively chirped bunch with the RF-Deflector turned off. (a) Initial energy spread without ($\sigma_E\approx 0.6$~MeV) and with plasma ($\sigma_E\approx 0.1$~MeV). In (b) the plasma density is $n_p\approx 1.8\times 10^{14}$~cm$^{-3}$.}
\label{espread}
\end{figure}

To measure the effect on the energy spectrum of the beam induced by the plasma, we transported the beam into the magnetic spectrometer downstream the capillary (with the RF-Deflector turned off) and made several measurements at different plasma densities. Once the $H_2$ is ionized it takes almost 10 $\mu s$ to recombine~\cite{filippi2016plasma}. During this time the plasma density is slowly decreases, thus by choosing the time-of arrival, by delaying the beam, we could choose the plasma density to interact with. Figure~\ref{EnergyNoPlasma} shows the unperturbed energy spectrum, when there is no plasma in the capillary. In this case the overall energy spread is $\sigma_E\approx 0.6$~MeV, similarly to Fig.~\ref{LPSbeam}. When the plasma is turned on and its density tuned to $n_p\approx 1.8\times 10^{14}$~cm$^{-3}$ (corresponding to a delay of the order of $4.5~\mu $s) we achieved the maximum reduction of the beam energy spread, down to $\sigma_E\approx 0.1$~MeV (see Fig.\ref{EnergyWithPlasma}).

\begin{figure}[h]
\centering
\includegraphics[width=0.9\linewidth]{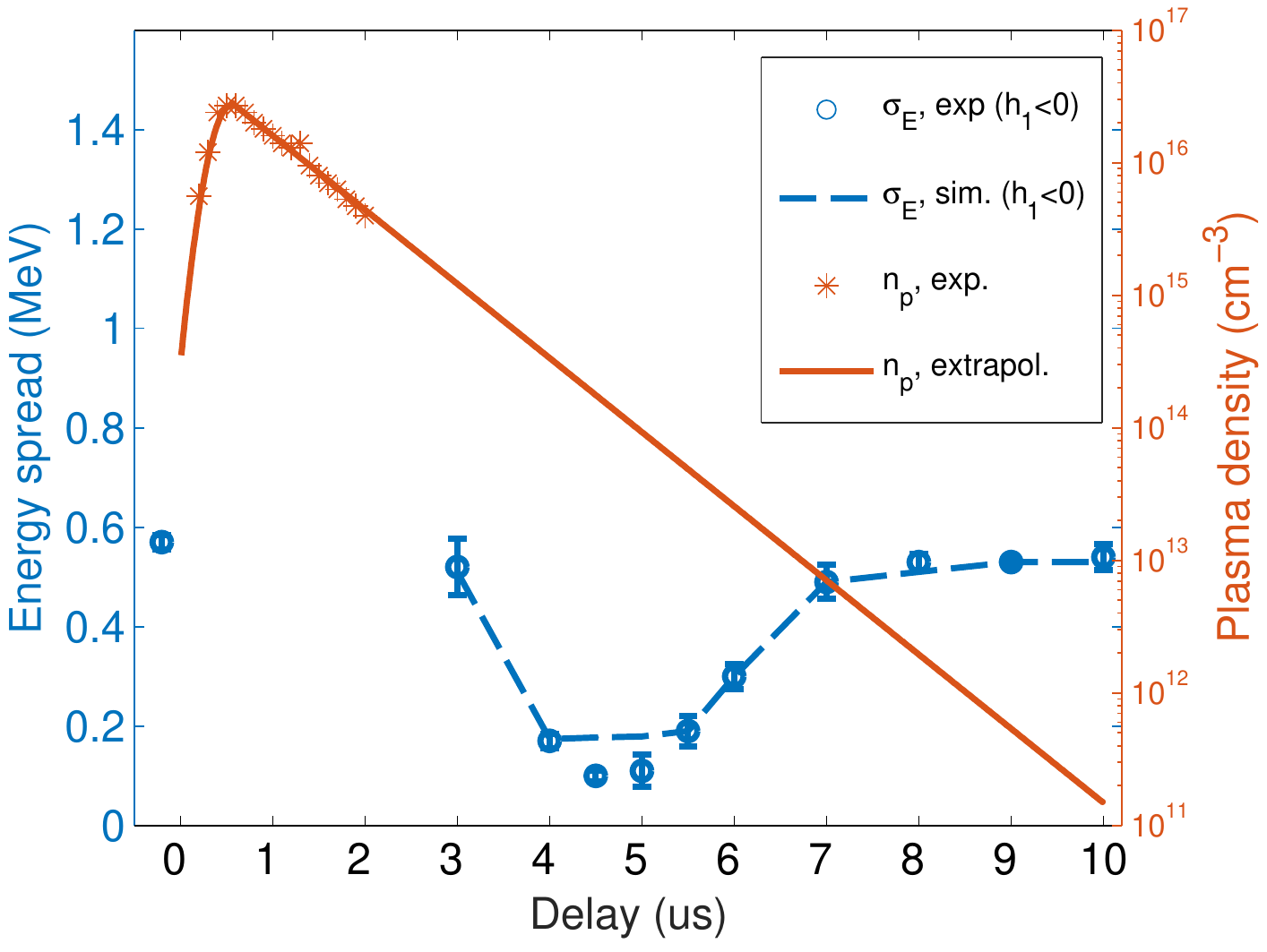}
\caption{Experimentally measured energy spread for the initially chirped electron bunch (blue circles) as a function of the delay with respect to the discharge trigger. The simulation results are depicted as dashed blue curve. The plasma density measured for several delays (red stars) and its extrapolated evolution (red line) are also reported.}
\label{PlasmaSpreadTotal}
\end{figure}

The evolution of the bunch energy spread for different plasma densities is shown in Fig.~\ref{PlasmaSpreadTotal}. Both quantities have been reported as a function of the delay of the discharge trigger. Being the Stark broadening diagnostics limited to the measurement of plasma densities above $\approx 10^{15}$~cm$^{-3}$ (red stars), for lower values the expected density can be extrapolated (red line) only theoretically~\cite{johnson1973ionization}. For the studied plasma densities the energy spread of the beam with initially negative chirp (see Figs.\ref{espread}) was decreasing, achieving its minimum at plasma density $n_p=1.8\times10^{14}$ (blue circles). The missing points on the energy spread curve correspond to a time when the discharge occurs and active lens effects are taking place~\cite{PhysRevLett.121.174801}.
%
%
%

The study on the manipulation of the LPS by the beam-driven plasma wakefields excited in a discharge-capillary structure is completed by analyzing the evolution of the negatively chirped beam configuration through the entire plasma channel.
The interaction is described by using a 2D plasma wakefield code~\cite{lu2005limits} that also takes into account the finite plasma radial extension, being confined within the capillary radius $R_c$~\cite{fang2014effect}. Following our previous studies in which we completely characterized the longitudinal plasma density profile along the capillary, here the channel is numerically computed by assuming a flat profile in the central part with decreasing exponential tails extending 1~cm outside the capillary~\cite{filippi2016spectroscopic,biagioni2016electron,filippi2018tapering}. The evolution of the bunch energy spread is shown in Fig.~\ref{PlasmaSpreadTotal} as dashed blue line. As input beam we have used results of start-to-end simulations of the SPARC\_LAB photo-injector by using the General Particle Tracer (GPT) code~\cite{gpt_web}, resulting in the LPS shown in Fig.~\ref{GPT_LPS_Wake}. The excited plasma wakefields act along the entire channel to decelerate the particles in the tail of the beam, resulting in a final energy spread of the order of 0.1~MeV for the negatively chirped beam (blue circles), in agreement with the experimental measurements reported in Fig.~\ref{PlasmaSpreadTotal}.
\begin{figure}[h]
\centering
\includegraphics[width=0.9\linewidth]{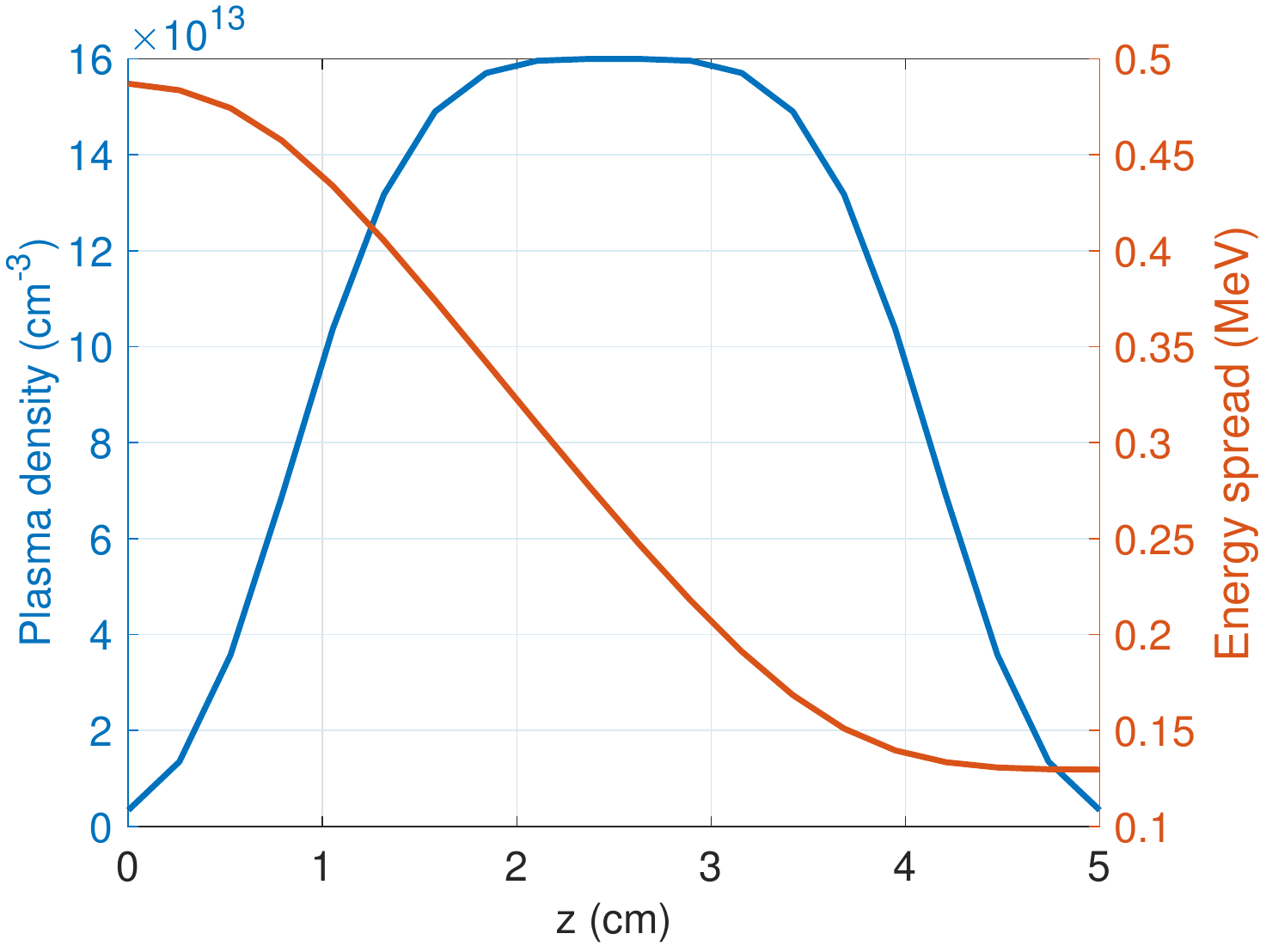}
\caption{Computed energy spread (red) of the chirped bunch along the plasma channel (blue). The plasma profile is calculated assuming the 3~cm-long capillary with 1~cm input and exit ramps.}
\label{SpreadEvol}
\end{figure}

Figure~\ref{SpreadEvol} shows the evolution of the reduction of the bunch energy spread along the plasma channel. Here we are referring to the plasma density that provides the best energy spread reduction. As one can see, most of the reduction happens inside the capillary, where the plasma density is larger. On the contrary, on the input and exit ramps the reduction is almost negligible due to the extremely low associated plasma densities.



In conclusion, we have demonstrated the use of plasma wakefield to manipulate the longitudinal phase-space of an electron beam. For this purpose we have conducted a proof-of-principle experiments where we completely characterized a plasma-based device consisting of a 3~cm–long capillary filled by H$_2$ gas. Our findings clearly proved that the large fields excited in a confined plasma can be used to tune the time-energy correlation of the particles according to the desired task.
We have shown that such a device is not only compact but also can be highly flexible. For the beam with negative chirp we demonstrated a possibility to completely remove energy chirp and reduced the total energy spread from 0.6 to 0.1 MeV (the level of uncorrelated energy spread of the SPARC photo-injector).

Several applications can benefit of such results. It represents, for example, an interesting tool for FEL facilities to imprint an energy-chirp in the beam and achieve shorter bunch lengths in a magnetic compressor.
The major advantage, however, is when employing this device downstream a plasma-based accelerator. It is well known that plasma-accelerated bunches have a large (negative) energy-chirp due to the larger fields experienced by the tail.
In this case a second plasma module, as the one we have reported, might be implemented in order to remove such a correlation and reduce the overall energy spread. Due to high flexibility of the plasma dechirper, by manipulating the parameters of the system (like plasma density) and parameters of the beam (changing its density with focusing), we can easily tune the system to exactly remove the given correlated energy spread.  It represents an essential feature in order to make the plasma-accelerated beams usable with conventional magnetic optics and in applications like Inverse Compton Scattering or FEL.

\begin{acknowledgments}
This work has been partially supported by the EU Commission in the Seventh Framework Program, Grant Agreement 312453-EuCARD-2 and the European Union Horizon 2020 research and innovation program, Grant Agreement No. 653782 (EuPRAXIA). The work of one of us (A.Z.) was partially supported by BSF foundation.
\end{acknowledgments}

\bibliography{biblio}

\begin{thebibliography}{46}%
\makeatletter
\providecommand \@ifxundefined [1]{%
 \@ifx{#1\undefined}
}%
\providecommand \@ifnum [1]{%
 \ifnum #1\expandafter \@firstoftwo
 \else \expandafter \@secondoftwo
 \fi
}%
\providecommand \@ifx [1]{%
 \ifx #1\expandafter \@firstoftwo
 \else \expandafter \@secondoftwo
 \fi
}%
\providecommand \natexlab [1]{#1}%
\providecommand \enquote  [1]{``#1''}%
\providecommand \bibnamefont  [1]{#1}%
\providecommand \bibfnamefont [1]{#1}%
\providecommand \citenamefont [1]{#1}%
\providecommand \href@noop [0]{\@secondoftwo}%
\providecommand \href [0]{\begingroup \@sanitize@url \@href}%
\providecommand \@href[1]{\@@startlink{#1}\@@href}%
\providecommand \@@href[1]{\endgroup#1\@@endlink}%
\providecommand \@sanitize@url [0]{\catcode `\\12\catcode `\$12\catcode
  `\&12\catcode `\#12\catcode `\^12\catcode `\_12\catcode `\%12\relax}%
\providecommand \@@startlink[1]{}%
\providecommand \@@endlink[0]{}%
\providecommand \url  [0]{\begingroup\@sanitize@url \@url }%
\providecommand \@url [1]{\endgroup\@href {#1}{\urlprefix }}%
\providecommand \urlprefix  [0]{URL }%
\providecommand \Eprint [0]{\href }%
\providecommand \doibase [0]{http://dx.doi.org/}%
\providecommand \selectlanguage [0]{\@gobble}%
\providecommand \bibinfo  [0]{\@secondoftwo}%
\providecommand \bibfield  [0]{\@secondoftwo}%
\providecommand \translation [1]{[#1]}%
\providecommand \BibitemOpen [0]{}%
\providecommand \bibitemStop [0]{}%
\providecommand \bibitemNoStop [0]{.\EOS\space}%
\providecommand \EOS [0]{\spacefactor3000\relax}%
\providecommand \BibitemShut  [1]{\csname bibitem#1\endcsname}%
\let\auto@bib@innerbib\@empty
\bibitem [{\citenamefont {Schoenlein}\ \emph {et~al.}(1996)\citenamefont
  {Schoenlein}, \citenamefont {Leemans}, \citenamefont {Chin},\ and\
  \citenamefont {Volfbeyn}}]{schoenlein1996femtosecond}%
  \BibitemOpen
  \bibfield  {author} {\bibinfo {author} {\bibfnamefont {R.~W.}\ \bibnamefont
  {Schoenlein}}, \bibinfo {author} {\bibfnamefont {W.}~\bibnamefont {Leemans}},
  \bibinfo {author} {\bibfnamefont {A.}~\bibnamefont {Chin}}, \ and\ \bibinfo
  {author} {\bibfnamefont {P.}~\bibnamefont {Volfbeyn}},\ }\href@noop {}
  {\bibfield  {journal} {\bibinfo  {journal} {Science}\ }\textbf {\bibinfo
  {volume} {274}},\ \bibinfo {pages} {236} (\bibinfo {year}
  {1996})}\BibitemShut {NoStop}%
\bibitem [{\citenamefont {Bacci}\ \emph {et~al.}(2013)\citenamefont {Bacci},
  \citenamefont {Alesini}, \citenamefont {Antici}, \citenamefont {Bellaveglia},
  \citenamefont {Boni}, \citenamefont {Chiadroni}, \citenamefont {Cianchi},
  \citenamefont {Curatolo}, \citenamefont {Di~Pirro}, \citenamefont {Esposito}
  \emph {et~al.}}]{bacci2013electron}%
  \BibitemOpen
  \bibfield  {author} {\bibinfo {author} {\bibfnamefont {A.}~\bibnamefont
  {Bacci}}, \bibinfo {author} {\bibfnamefont {D.}~\bibnamefont {Alesini}},
  \bibinfo {author} {\bibfnamefont {P.}~\bibnamefont {Antici}}, \bibinfo
  {author} {\bibfnamefont {M.}~\bibnamefont {Bellaveglia}}, \bibinfo {author}
  {\bibfnamefont {R.}~\bibnamefont {Boni}}, \bibinfo {author} {\bibfnamefont
  {E.}~\bibnamefont {Chiadroni}}, \bibinfo {author} {\bibfnamefont
  {A.}~\bibnamefont {Cianchi}}, \bibinfo {author} {\bibfnamefont
  {C.}~\bibnamefont {Curatolo}}, \bibinfo {author} {\bibfnamefont
  {G.}~\bibnamefont {Di~Pirro}}, \bibinfo {author} {\bibfnamefont
  {A.}~\bibnamefont {Esposito}},  \emph {et~al.},\ }\href@noop {} {\bibfield
  {journal} {\bibinfo  {journal} {Journal of Applied Physics}\ }\textbf
  {\bibinfo {volume} {113}},\ \bibinfo {pages} {194508} (\bibinfo {year}
  {2013})}\BibitemShut {NoStop}%
\bibitem [{\citenamefont {Chiadroni}\ \emph
  {et~al.}(2013{\natexlab{a}})\citenamefont {Chiadroni}, \citenamefont
  {Bellaveglia}, \citenamefont {Calvani}, \citenamefont {Castellano},
  \citenamefont {Catani}, \citenamefont {Cianchi}, \citenamefont {Pirro},
  \citenamefont {Ferrario}, \citenamefont {Gatti}, \citenamefont {Limaj},
  \citenamefont {Lupi}, \citenamefont {Marchetti}, \citenamefont {Mostacci},
  \citenamefont {Pace}, \citenamefont {Palumbo}, \citenamefont {Ronsivalle},
  \citenamefont {Pompili},\ and\ \citenamefont
  {Vaccarezza}}]{chiadroni:022703}%
  \BibitemOpen
  \bibfield  {author} {\bibinfo {author} {\bibfnamefont {E.}~\bibnamefont
  {Chiadroni}}, \bibinfo {author} {\bibfnamefont {M.}~\bibnamefont
  {Bellaveglia}}, \bibinfo {author} {\bibfnamefont {P.}~\bibnamefont
  {Calvani}}, \bibinfo {author} {\bibfnamefont {M.}~\bibnamefont {Castellano}},
  \bibinfo {author} {\bibfnamefont {L.}~\bibnamefont {Catani}}, \bibinfo
  {author} {\bibfnamefont {A.}~\bibnamefont {Cianchi}}, \bibinfo {author}
  {\bibfnamefont {G.~D.}\ \bibnamefont {Pirro}}, \bibinfo {author}
  {\bibfnamefont {M.}~\bibnamefont {Ferrario}}, \bibinfo {author}
  {\bibfnamefont {G.}~\bibnamefont {Gatti}}, \bibinfo {author} {\bibfnamefont
  {O.}~\bibnamefont {Limaj}}, \bibinfo {author} {\bibfnamefont
  {S.}~\bibnamefont {Lupi}}, \bibinfo {author} {\bibfnamefont {B.}~\bibnamefont
  {Marchetti}}, \bibinfo {author} {\bibfnamefont {A.}~\bibnamefont {Mostacci}},
  \bibinfo {author} {\bibfnamefont {E.}~\bibnamefont {Pace}}, \bibinfo {author}
  {\bibfnamefont {L.}~\bibnamefont {Palumbo}}, \bibinfo {author} {\bibfnamefont
  {C.}~\bibnamefont {Ronsivalle}}, \bibinfo {author} {\bibfnamefont
  {R.}~\bibnamefont {Pompili}}, \ and\ \bibinfo {author} {\bibfnamefont
  {C.}~\bibnamefont {Vaccarezza}},\ }\href {\doibase 10.1063/1.4790429}
  {\bibfield  {journal} {\bibinfo  {journal} {Review of Scientific
  Instruments}\ }\textbf {\bibinfo {volume} {84}},\ \bibinfo {eid} {022703}
  (\bibinfo {year} {2013}{\natexlab{a}})}\BibitemShut {NoStop}%
\bibitem [{\citenamefont {Giorgianni}\ \emph {et~al.}(2016)\citenamefont
  {Giorgianni}, \citenamefont {Chiadroni}, \citenamefont {Rovere},
  \citenamefont {Cestelli-Guidi}, \citenamefont {Perucchi}, \citenamefont
  {Bellaveglia}, \citenamefont {Castellano}, \citenamefont {Di~Giovenale},
  \citenamefont {Di~Pirro}, \citenamefont {Ferrario} \emph
  {et~al.}}]{giorgianni2016strong}%
  \BibitemOpen
  \bibfield  {author} {\bibinfo {author} {\bibfnamefont {F.}~\bibnamefont
  {Giorgianni}}, \bibinfo {author} {\bibfnamefont {E.}~\bibnamefont
  {Chiadroni}}, \bibinfo {author} {\bibfnamefont {A.}~\bibnamefont {Rovere}},
  \bibinfo {author} {\bibfnamefont {M.}~\bibnamefont {Cestelli-Guidi}},
  \bibinfo {author} {\bibfnamefont {A.}~\bibnamefont {Perucchi}}, \bibinfo
  {author} {\bibfnamefont {M.}~\bibnamefont {Bellaveglia}}, \bibinfo {author}
  {\bibfnamefont {M.}~\bibnamefont {Castellano}}, \bibinfo {author}
  {\bibfnamefont {D.}~\bibnamefont {Di~Giovenale}}, \bibinfo {author}
  {\bibfnamefont {G.}~\bibnamefont {Di~Pirro}}, \bibinfo {author}
  {\bibfnamefont {M.}~\bibnamefont {Ferrario}},  \emph {et~al.},\ }\href@noop
  {} {\bibfield  {journal} {\bibinfo  {journal} {Nature communications}\
  }\textbf {\bibinfo {volume} {7}} (\bibinfo {year} {2016})}\BibitemShut
  {NoStop}%
\bibitem [{\citenamefont {Ackermann}\ \emph {et~al.}(2007)\citenamefont
  {Ackermann}, \citenamefont {Asova}, \citenamefont {Ayvazyan}, \citenamefont
  {Azima}, \citenamefont {Baboi}, \citenamefont {Balandin}, \citenamefont
  {Beutner}, \citenamefont {Brandt}, \citenamefont {Bolzmann} \emph
  {et~al.}}]{ackermann2007operation}%
  \BibitemOpen
  \bibfield  {author} {\bibinfo {author} {\bibfnamefont {W.}~\bibnamefont
  {Ackermann}}, \bibinfo {author} {\bibfnamefont {G.}~\bibnamefont {Asova}},
  \bibinfo {author} {\bibfnamefont {V.}~\bibnamefont {Ayvazyan}}, \bibinfo
  {author} {\bibfnamefont {A.}~\bibnamefont {Azima}}, \bibinfo {author}
  {\bibfnamefont {J.}~\bibnamefont {Baboi}}, \bibinfo {author} {\bibfnamefont
  {V.}~\bibnamefont {Balandin}}, \bibinfo {author} {\bibfnamefont
  {B.}~\bibnamefont {Beutner}}, \bibinfo {author} {\bibfnamefont
  {A.}~\bibnamefont {Brandt}}, \bibinfo {author} {\bibfnamefont
  {A.}~\bibnamefont {Bolzmann}},  \emph {et~al.},\ }\href@noop {} {\bibfield
  {journal} {\bibinfo  {journal} {Nature photonics}\ }\textbf {\bibinfo
  {volume} {1}},\ \bibinfo {pages} {336} (\bibinfo {year} {2007})}\BibitemShut
  {NoStop}%
\bibitem [{\citenamefont {Emma}\ \emph {et~al.}(2010)\citenamefont {Emma},
  \citenamefont {Akre}, \citenamefont {Arthur}, \citenamefont {Bionta},
  \citenamefont {Bostedt}, \citenamefont {Bozek}, \citenamefont {Brachmann},
  \citenamefont {Bucksbaum}, \citenamefont {Coffee}, \citenamefont {Decker}
  \emph {et~al.}}]{emma2010first}%
  \BibitemOpen
  \bibfield  {author} {\bibinfo {author} {\bibfnamefont {P.}~\bibnamefont
  {Emma}}, \bibinfo {author} {\bibfnamefont {R.}~\bibnamefont {Akre}}, \bibinfo
  {author} {\bibfnamefont {J.}~\bibnamefont {Arthur}}, \bibinfo {author}
  {\bibfnamefont {R.}~\bibnamefont {Bionta}}, \bibinfo {author} {\bibfnamefont
  {C.}~\bibnamefont {Bostedt}}, \bibinfo {author} {\bibfnamefont
  {J.}~\bibnamefont {Bozek}}, \bibinfo {author} {\bibfnamefont
  {A.}~\bibnamefont {Brachmann}}, \bibinfo {author} {\bibfnamefont
  {P.}~\bibnamefont {Bucksbaum}}, \bibinfo {author} {\bibfnamefont
  {R.}~\bibnamefont {Coffee}}, \bibinfo {author} {\bibfnamefont {F.-J.}\
  \bibnamefont {Decker}},  \emph {et~al.},\ }\href@noop {} {\bibfield
  {journal} {\bibinfo  {journal} {nature photonics}\ }\textbf {\bibinfo
  {volume} {4}},\ \bibinfo {pages} {641} (\bibinfo {year} {2010})}\BibitemShut
  {NoStop}%
\bibitem [{\citenamefont {Allaria}\ \emph {et~al.}(2012)\citenamefont
  {Allaria}, \citenamefont {Appio}, \citenamefont {Badano}, \citenamefont
  {Barletta}, \citenamefont {Bassanese}, \citenamefont {Biedron}, \citenamefont
  {Borga}, \citenamefont {Busetto}, \citenamefont {Castronovo}, \citenamefont
  {Cinquegrana} \emph {et~al.}}]{allaria2012highly}%
  \BibitemOpen
  \bibfield  {author} {\bibinfo {author} {\bibfnamefont {E.}~\bibnamefont
  {Allaria}}, \bibinfo {author} {\bibfnamefont {R.}~\bibnamefont {Appio}},
  \bibinfo {author} {\bibfnamefont {L.}~\bibnamefont {Badano}}, \bibinfo
  {author} {\bibfnamefont {W.}~\bibnamefont {Barletta}}, \bibinfo {author}
  {\bibfnamefont {S.}~\bibnamefont {Bassanese}}, \bibinfo {author}
  {\bibfnamefont {S.}~\bibnamefont {Biedron}}, \bibinfo {author} {\bibfnamefont
  {A.}~\bibnamefont {Borga}}, \bibinfo {author} {\bibfnamefont
  {E.}~\bibnamefont {Busetto}}, \bibinfo {author} {\bibfnamefont
  {D.}~\bibnamefont {Castronovo}}, \bibinfo {author} {\bibfnamefont
  {P.}~\bibnamefont {Cinquegrana}},  \emph {et~al.},\ }\href@noop {} {\bibfield
   {journal} {\bibinfo  {journal} {Nature Photonics}\ }\textbf {\bibinfo
  {volume} {6}},\ \bibinfo {pages} {699} (\bibinfo {year} {2012})}\BibitemShut
  {NoStop}%
\bibitem [{\citenamefont {Petrillo}\ \emph {et~al.}(2013)\citenamefont
  {Petrillo}, \citenamefont {Anania}, \citenamefont {Artioli}, \citenamefont
  {Bacci}, \citenamefont {Bellaveglia}, \citenamefont {Chiadroni},
  \citenamefont {Cianchi}, \citenamefont {Ciocci}, \citenamefont {Dattoli},
  \citenamefont {Di~Giovenale} \emph {et~al.}}]{petrillo2013observation}%
  \BibitemOpen
  \bibfield  {author} {\bibinfo {author} {\bibfnamefont {V.}~\bibnamefont
  {Petrillo}}, \bibinfo {author} {\bibfnamefont {M.}~\bibnamefont {Anania}},
  \bibinfo {author} {\bibfnamefont {M.}~\bibnamefont {Artioli}}, \bibinfo
  {author} {\bibfnamefont {A.}~\bibnamefont {Bacci}}, \bibinfo {author}
  {\bibfnamefont {M.}~\bibnamefont {Bellaveglia}}, \bibinfo {author}
  {\bibfnamefont {E.}~\bibnamefont {Chiadroni}}, \bibinfo {author}
  {\bibfnamefont {A.}~\bibnamefont {Cianchi}}, \bibinfo {author} {\bibfnamefont
  {F.}~\bibnamefont {Ciocci}}, \bibinfo {author} {\bibfnamefont
  {G.}~\bibnamefont {Dattoli}}, \bibinfo {author} {\bibfnamefont
  {D.}~\bibnamefont {Di~Giovenale}},  \emph {et~al.},\ }\href@noop {}
  {\bibfield  {journal} {\bibinfo  {journal} {Physical Review Letters}\
  }\textbf {\bibinfo {volume} {111}},\ \bibinfo {pages} {114802} (\bibinfo
  {year} {2013})}\BibitemShut {NoStop}%
\bibitem [{\citenamefont {{Tajima}}\ and\ \citenamefont
  {{Dawson}}(1979)}]{1979PhRvL..43..267T}%
  \BibitemOpen
  \bibfield  {author} {\bibinfo {author} {\bibfnamefont {T.}~\bibnamefont
  {{Tajima}}}\ and\ \bibinfo {author} {\bibfnamefont {J.~M.}\ \bibnamefont
  {{Dawson}}},\ }\href {\doibase 10.1103/PhysRevLett.43.267} {\bibfield
  {journal} {\bibinfo  {journal} {Physical Review Letters}\ }\textbf {\bibinfo
  {volume} {43}},\ \bibinfo {pages} {267} (\bibinfo {year} {1979})}\BibitemShut
  {NoStop}%
\bibitem [{\citenamefont {{Rosenzweig}}(1987)}]{1987PhRvL..58..555R}%
  \BibitemOpen
  \bibfield  {author} {\bibinfo {author} {\bibfnamefont {J.~B.}\ \bibnamefont
  {{Rosenzweig}}},\ }\href {\doibase 10.1103/PhysRevLett.58.555} {\bibfield
  {journal} {\bibinfo  {journal} {Physical Review Letters}\ }\textbf {\bibinfo
  {volume} {58}},\ \bibinfo {pages} {555} (\bibinfo {year} {1987})}\BibitemShut
  {NoStop}%
\bibitem [{\citenamefont {Rosenzweig}\ \emph {et~al.}(1991)\citenamefont
  {Rosenzweig}, \citenamefont {Breizman}, \citenamefont {Katsouleas},\ and\
  \citenamefont {Su}}]{rosenzweig1991acceleration}%
  \BibitemOpen
  \bibfield  {author} {\bibinfo {author} {\bibfnamefont {J.}~\bibnamefont
  {Rosenzweig}}, \bibinfo {author} {\bibfnamefont {B.}~\bibnamefont
  {Breizman}}, \bibinfo {author} {\bibfnamefont {T.}~\bibnamefont
  {Katsouleas}}, \ and\ \bibinfo {author} {\bibfnamefont {J.}~\bibnamefont
  {Su}},\ }\href@noop {} {\bibfield  {journal} {\bibinfo  {journal} {Physical
  Review A}\ }\textbf {\bibinfo {volume} {44}},\ \bibinfo {pages} {R6189}
  (\bibinfo {year} {1991})}\BibitemShut {NoStop}%
\bibitem [{\citenamefont {Litos}\ \emph {et~al.}(2014)\citenamefont {Litos},
  \citenamefont {Adli}, \citenamefont {An}, \citenamefont {Clarke},
  \citenamefont {Clayton}, \citenamefont {Corde}, \citenamefont {Delahaye},
  \citenamefont {England}, \citenamefont {Fisher}, \citenamefont {Frederico}
  \emph {et~al.}}]{litos2014high}%
  \BibitemOpen
  \bibfield  {author} {\bibinfo {author} {\bibfnamefont {M.}~\bibnamefont
  {Litos}}, \bibinfo {author} {\bibfnamefont {E.}~\bibnamefont {Adli}},
  \bibinfo {author} {\bibfnamefont {W.}~\bibnamefont {An}}, \bibinfo {author}
  {\bibfnamefont {C.}~\bibnamefont {Clarke}}, \bibinfo {author} {\bibfnamefont
  {C.}~\bibnamefont {Clayton}}, \bibinfo {author} {\bibfnamefont
  {S.}~\bibnamefont {Corde}}, \bibinfo {author} {\bibfnamefont
  {J.}~\bibnamefont {Delahaye}}, \bibinfo {author} {\bibfnamefont
  {R.}~\bibnamefont {England}}, \bibinfo {author} {\bibfnamefont
  {A.}~\bibnamefont {Fisher}}, \bibinfo {author} {\bibfnamefont
  {J.}~\bibnamefont {Frederico}},  \emph {et~al.},\ }\href@noop {} {\bibfield
  {journal} {\bibinfo  {journal} {Nature}\ }\textbf {\bibinfo {volume} {515}},\
  \bibinfo {pages} {92} (\bibinfo {year} {2014})}\BibitemShut {NoStop}%
\bibitem [{\citenamefont {Giannessi}\ \emph {et~al.}(2011)\citenamefont
  {Giannessi}, \citenamefont {Bacci}, \citenamefont {Bellaveglia},
  \citenamefont {Briquez}, \citenamefont {Castellano}, \citenamefont
  {Chiadroni}, \citenamefont {Cianchi}, \citenamefont {Ciocci}, \citenamefont
  {Couprie}, \citenamefont {Cultrera} \emph {et~al.}}]{giannessi2011self}%
  \BibitemOpen
  \bibfield  {author} {\bibinfo {author} {\bibfnamefont {L.}~\bibnamefont
  {Giannessi}}, \bibinfo {author} {\bibfnamefont {A.}~\bibnamefont {Bacci}},
  \bibinfo {author} {\bibfnamefont {M.}~\bibnamefont {Bellaveglia}}, \bibinfo
  {author} {\bibfnamefont {F.}~\bibnamefont {Briquez}}, \bibinfo {author}
  {\bibfnamefont {M.}~\bibnamefont {Castellano}}, \bibinfo {author}
  {\bibfnamefont {E.}~\bibnamefont {Chiadroni}}, \bibinfo {author}
  {\bibfnamefont {A.}~\bibnamefont {Cianchi}}, \bibinfo {author} {\bibfnamefont
  {F.}~\bibnamefont {Ciocci}}, \bibinfo {author} {\bibfnamefont
  {M.}~\bibnamefont {Couprie}}, \bibinfo {author} {\bibfnamefont
  {L.}~\bibnamefont {Cultrera}},  \emph {et~al.},\ }\href@noop {} {\bibfield
  {journal} {\bibinfo  {journal} {Physical review letters}\ }\textbf {\bibinfo
  {volume} {106}},\ \bibinfo {pages} {144801} (\bibinfo {year}
  {2011})}\BibitemShut {NoStop}%
\bibitem [{\citenamefont {{Blumenfeld}}\ \emph {et~al.}(2007)\citenamefont
  {{Blumenfeld}}, \citenamefont {{Clayton}}, \citenamefont {{Decker}},
  \citenamefont {{Hogan}}, \citenamefont {{Huang}}, \citenamefont
  {{Ischebeck}}, \citenamefont {{Iverson}}, \citenamefont {{Joshi}},
  \citenamefont {{Katsouleas}}, \citenamefont {{Kirby}}, \citenamefont {{Lu}},
  \citenamefont {{Marsh}}, \citenamefont {{Mori}}, \citenamefont {{Muggli}},
  \citenamefont {{Oz}}, \citenamefont {{Siemann}}, \citenamefont {{Walz}},\
  and\ \citenamefont {{Zhou}}}]{2007Natur.445..741B}%
  \BibitemOpen
  \bibfield  {author} {\bibinfo {author} {\bibfnamefont {I.}~\bibnamefont
  {{Blumenfeld}}}, \bibinfo {author} {\bibfnamefont {C.~E.}\ \bibnamefont
  {{Clayton}}}, \bibinfo {author} {\bibfnamefont {F.-J.}\ \bibnamefont
  {{Decker}}}, \bibinfo {author} {\bibfnamefont {M.~J.}\ \bibnamefont
  {{Hogan}}}, \bibinfo {author} {\bibfnamefont {C.}~\bibnamefont {{Huang}}},
  \bibinfo {author} {\bibfnamefont {R.}~\bibnamefont {{Ischebeck}}}, \bibinfo
  {author} {\bibfnamefont {R.}~\bibnamefont {{Iverson}}}, \bibinfo {author}
  {\bibfnamefont {C.}~\bibnamefont {{Joshi}}}, \bibinfo {author} {\bibfnamefont
  {T.}~\bibnamefont {{Katsouleas}}}, \bibinfo {author} {\bibfnamefont
  {N.}~\bibnamefont {{Kirby}}}, \bibinfo {author} {\bibfnamefont
  {W.}~\bibnamefont {{Lu}}}, \bibinfo {author} {\bibfnamefont {K.~A.}\
  \bibnamefont {{Marsh}}}, \bibinfo {author} {\bibfnamefont {W.~B.}\
  \bibnamefont {{Mori}}}, \bibinfo {author} {\bibfnamefont {P.}~\bibnamefont
  {{Muggli}}}, \bibinfo {author} {\bibfnamefont {E.}~\bibnamefont {{Oz}}},
  \bibinfo {author} {\bibfnamefont {R.~H.}\ \bibnamefont {{Siemann}}}, \bibinfo
  {author} {\bibfnamefont {D.}~\bibnamefont {{Walz}}}, \ and\ \bibinfo {author}
  {\bibfnamefont {M.}~\bibnamefont {{Zhou}}},\ }\href {\doibase
  10.1038/nature05538} {\bibfield  {journal} {\bibinfo  {journal} {Nature}\
  }\textbf {\bibinfo {volume} {445}},\ \bibinfo {pages} {741} (\bibinfo {year}
  {2007})}\BibitemShut {NoStop}%
\bibitem [{\citenamefont {Leemans}\ \emph {et~al.}(2006)\citenamefont
  {Leemans}, \citenamefont {Nagler}, \citenamefont {Gonsalves}, \citenamefont
  {T{\'o}th}, \citenamefont {Nakamura}, \citenamefont {Geddes}, \citenamefont
  {Esarey}, \citenamefont {Schroeder},\ and\ \citenamefont
  {Hooker}}]{leemans2006gev}%
  \BibitemOpen
  \bibfield  {author} {\bibinfo {author} {\bibfnamefont {W.}~\bibnamefont
  {Leemans}}, \bibinfo {author} {\bibfnamefont {B.}~\bibnamefont {Nagler}},
  \bibinfo {author} {\bibfnamefont {A.}~\bibnamefont {Gonsalves}}, \bibinfo
  {author} {\bibfnamefont {C.}~\bibnamefont {T{\'o}th}}, \bibinfo {author}
  {\bibfnamefont {K.}~\bibnamefont {Nakamura}}, \bibinfo {author}
  {\bibfnamefont {C.}~\bibnamefont {Geddes}}, \bibinfo {author} {\bibfnamefont
  {E.}~\bibnamefont {Esarey}}, \bibinfo {author} {\bibfnamefont
  {C.}~\bibnamefont {Schroeder}}, \ and\ \bibinfo {author} {\bibfnamefont
  {S.}~\bibnamefont {Hooker}},\ }\href@noop {} {\bibfield  {journal} {\bibinfo
  {journal} {Nature physics}\ }\textbf {\bibinfo {volume} {2}},\ \bibinfo
  {pages} {696} (\bibinfo {year} {2006})}\BibitemShut {NoStop}%
\bibitem [{\citenamefont {Faure}\ \emph {et~al.}(2006)\citenamefont {Faure},
  \citenamefont {Rechatin}, \citenamefont {Norlin}, \citenamefont {Lifschitz},
  \citenamefont {Glinec},\ and\ \citenamefont {Malka}}]{faure2006controlled}%
  \BibitemOpen
  \bibfield  {author} {\bibinfo {author} {\bibfnamefont {J.}~\bibnamefont
  {Faure}}, \bibinfo {author} {\bibfnamefont {C.}~\bibnamefont {Rechatin}},
  \bibinfo {author} {\bibfnamefont {A.}~\bibnamefont {Norlin}}, \bibinfo
  {author} {\bibfnamefont {A.}~\bibnamefont {Lifschitz}}, \bibinfo {author}
  {\bibfnamefont {Y.}~\bibnamefont {Glinec}}, \ and\ \bibinfo {author}
  {\bibfnamefont {V.}~\bibnamefont {Malka}},\ }\href@noop {} {\bibfield
  {journal} {\bibinfo  {journal} {Nature}\ }\textbf {\bibinfo {volume} {444}},\
  \bibinfo {pages} {737} (\bibinfo {year} {2006})}\BibitemShut {NoStop}%
\bibitem [{\citenamefont {Piot}\ \emph {et~al.}(2003)\citenamefont {Piot},
  \citenamefont {Douglas},\ and\ \citenamefont
  {Krafft}}]{piot2003longitudinal}%
  \BibitemOpen
  \bibfield  {author} {\bibinfo {author} {\bibfnamefont {P.}~\bibnamefont
  {Piot}}, \bibinfo {author} {\bibfnamefont {D.}~\bibnamefont {Douglas}}, \
  and\ \bibinfo {author} {\bibfnamefont {G.}~\bibnamefont {Krafft}},\
  }\href@noop {} {\bibfield  {journal} {\bibinfo  {journal} {Physical Review
  Special Topics-Accelerators and Beams}\ }\textbf {\bibinfo {volume} {6}},\
  \bibinfo {pages} {030702} (\bibinfo {year} {2003})}\BibitemShut {NoStop}%
\bibitem [{\citenamefont {England}\ \emph {et~al.}(2005)\citenamefont
  {England}, \citenamefont {Rosenzweig}, \citenamefont {Andonian},
  \citenamefont {Musumeci}, \citenamefont {Travish},\ and\ \citenamefont
  {Yoder}}]{england2005sextupole}%
  \BibitemOpen
  \bibfield  {author} {\bibinfo {author} {\bibfnamefont {R.}~\bibnamefont
  {England}}, \bibinfo {author} {\bibfnamefont {J.}~\bibnamefont {Rosenzweig}},
  \bibinfo {author} {\bibfnamefont {G.}~\bibnamefont {Andonian}}, \bibinfo
  {author} {\bibfnamefont {P.}~\bibnamefont {Musumeci}}, \bibinfo {author}
  {\bibfnamefont {G.}~\bibnamefont {Travish}}, \ and\ \bibinfo {author}
  {\bibfnamefont {R.}~\bibnamefont {Yoder}},\ }\href@noop {} {\bibfield
  {journal} {\bibinfo  {journal} {Physical Review Special Topics-Accelerators
  and Beams}\ }\textbf {\bibinfo {volume} {8}},\ \bibinfo {pages} {012801}
  (\bibinfo {year} {2005})}\BibitemShut {NoStop}%
\bibitem [{\citenamefont {Antipov}\ \emph {et~al.}(2014)\citenamefont
  {Antipov}, \citenamefont {Baturin}, \citenamefont {Jing}, \citenamefont
  {Fedurin}, \citenamefont {Kanareykin}, \citenamefont {Swinson}, \citenamefont
  {Schoessow}, \citenamefont {Gai},\ and\ \citenamefont
  {Zholents}}]{antipov2014experimental}%
  \BibitemOpen
  \bibfield  {author} {\bibinfo {author} {\bibfnamefont {S.}~\bibnamefont
  {Antipov}}, \bibinfo {author} {\bibfnamefont {S.}~\bibnamefont {Baturin}},
  \bibinfo {author} {\bibfnamefont {C.}~\bibnamefont {Jing}}, \bibinfo {author}
  {\bibfnamefont {M.}~\bibnamefont {Fedurin}}, \bibinfo {author} {\bibfnamefont
  {A.}~\bibnamefont {Kanareykin}}, \bibinfo {author} {\bibfnamefont
  {C.}~\bibnamefont {Swinson}}, \bibinfo {author} {\bibfnamefont
  {P.}~\bibnamefont {Schoessow}}, \bibinfo {author} {\bibfnamefont
  {W.}~\bibnamefont {Gai}}, \ and\ \bibinfo {author} {\bibfnamefont
  {A.}~\bibnamefont {Zholents}},\ }\href@noop {} {\bibfield  {journal}
  {\bibinfo  {journal} {Physical review letters}\ }\textbf {\bibinfo {volume}
  {112}},\ \bibinfo {pages} {114801} (\bibinfo {year} {2014})}\BibitemShut
  {NoStop}%
\bibitem [{\citenamefont {Bettoni}\ \emph {et~al.}(2016)\citenamefont
  {Bettoni}, \citenamefont {Craievich}, \citenamefont {Lutman},\ and\
  \citenamefont {Pedrozzi}}]{bettoni2016temporal}%
  \BibitemOpen
  \bibfield  {author} {\bibinfo {author} {\bibfnamefont {S.}~\bibnamefont
  {Bettoni}}, \bibinfo {author} {\bibfnamefont {P.}~\bibnamefont {Craievich}},
  \bibinfo {author} {\bibfnamefont {A.}~\bibnamefont {Lutman}}, \ and\ \bibinfo
  {author} {\bibfnamefont {M.}~\bibnamefont {Pedrozzi}},\ }\href@noop {}
  {\bibfield  {journal} {\bibinfo  {journal} {Physical Review Accelerators and
  Beams}\ }\textbf {\bibinfo {volume} {19}},\ \bibinfo {pages} {021304}
  (\bibinfo {year} {2016})}\BibitemShut {NoStop}%
\bibitem [{\citenamefont {Penco}\ \emph {et~al.}(2017)\citenamefont {Penco},
  \citenamefont {Allaria}, \citenamefont {Cudin}, \citenamefont {Di~Mitri},
  \citenamefont {Gauthier}, \citenamefont {Spampinati}, \citenamefont
  {Trov{\'o}}, \citenamefont {Giannessi}, \citenamefont {Roussel},
  \citenamefont {Bettoni} \emph {et~al.}}]{penco2017passive}%
  \BibitemOpen
  \bibfield  {author} {\bibinfo {author} {\bibfnamefont {G.}~\bibnamefont
  {Penco}}, \bibinfo {author} {\bibfnamefont {E.}~\bibnamefont {Allaria}},
  \bibinfo {author} {\bibfnamefont {I.}~\bibnamefont {Cudin}}, \bibinfo
  {author} {\bibfnamefont {S.}~\bibnamefont {Di~Mitri}}, \bibinfo {author}
  {\bibfnamefont {D.}~\bibnamefont {Gauthier}}, \bibinfo {author}
  {\bibfnamefont {S.}~\bibnamefont {Spampinati}}, \bibinfo {author}
  {\bibfnamefont {M.}~\bibnamefont {Trov{\'o}}}, \bibinfo {author}
  {\bibfnamefont {L.}~\bibnamefont {Giannessi}}, \bibinfo {author}
  {\bibfnamefont {E.}~\bibnamefont {Roussel}}, \bibinfo {author} {\bibfnamefont
  {S.}~\bibnamefont {Bettoni}},  \emph {et~al.},\ }\href@noop {} {\bibfield
  {journal} {\bibinfo  {journal} {Physical review letters}\ }\textbf {\bibinfo
  {volume} {119}},\ \bibinfo {pages} {184802} (\bibinfo {year}
  {2017})}\BibitemShut {NoStop}%
\bibitem [{\citenamefont {Wu}\ \emph {et~al.}(2017)\citenamefont {Wu},
  \citenamefont {Du}, \citenamefont {Zhang}, \citenamefont {Zhou},
  \citenamefont {Cheng}, \citenamefont {Zhou}, \citenamefont {Hua},
  \citenamefont {Pai},\ and\ \citenamefont {Lu}}]{wupreliminary}%
  \BibitemOpen
  \bibfield  {author} {\bibinfo {author} {\bibfnamefont {Y.}~\bibnamefont
  {Wu}}, \bibinfo {author} {\bibfnamefont {Y.}~\bibnamefont {Du}}, \bibinfo
  {author} {\bibfnamefont {J.}~\bibnamefont {Zhang}}, \bibinfo {author}
  {\bibfnamefont {Z.}~\bibnamefont {Zhou}}, \bibinfo {author} {\bibfnamefont
  {Z.}~\bibnamefont {Cheng}}, \bibinfo {author} {\bibfnamefont
  {S.}~\bibnamefont {Zhou}}, \bibinfo {author} {\bibfnamefont {J.}~\bibnamefont
  {Hua}}, \bibinfo {author} {\bibfnamefont {C.}~\bibnamefont {Pai}}, \ and\
  \bibinfo {author} {\bibfnamefont {W.}~\bibnamefont {Lu}},\ }\href@noop {}
  {\bibfield  {journal} {\bibinfo  {journal} {Proc. IPAC2017}\ } (\bibinfo
  {year} {2017})}\BibitemShut {NoStop}%
\bibitem [{\citenamefont {D’Arcy}\ \emph {et~al.}(2019)\citenamefont
  {D’Arcy}, \citenamefont {Wesch}, \citenamefont {Aschikhin}, \citenamefont
  {Bohlen}, \citenamefont {Behrens}, \citenamefont {Garland}, \citenamefont
  {Goldberg}, \citenamefont {Gonzalez}, \citenamefont {Knetsch}, \citenamefont
  {Libov} \emph {et~al.}}]{d2019tunable}%
  \BibitemOpen
  \bibfield  {author} {\bibinfo {author} {\bibfnamefont {R.}~\bibnamefont
  {D’Arcy}}, \bibinfo {author} {\bibfnamefont {S.}~\bibnamefont {Wesch}},
  \bibinfo {author} {\bibfnamefont {A.}~\bibnamefont {Aschikhin}}, \bibinfo
  {author} {\bibfnamefont {S.}~\bibnamefont {Bohlen}}, \bibinfo {author}
  {\bibfnamefont {C.}~\bibnamefont {Behrens}}, \bibinfo {author} {\bibfnamefont
  {M.}~\bibnamefont {Garland}}, \bibinfo {author} {\bibfnamefont
  {L.}~\bibnamefont {Goldberg}}, \bibinfo {author} {\bibfnamefont
  {P.}~\bibnamefont {Gonzalez}}, \bibinfo {author} {\bibfnamefont
  {A.}~\bibnamefont {Knetsch}}, \bibinfo {author} {\bibfnamefont
  {V.}~\bibnamefont {Libov}},  \emph {et~al.},\ }\href@noop {} {\bibfield
  {journal} {\bibinfo  {journal} {Physical Review Letters}\ }\textbf {\bibinfo
  {volume} {122}},\ \bibinfo {pages} {034801} (\bibinfo {year}
  {2019})}\BibitemShut {NoStop}%
\bibitem [{\citenamefont {Fang}\ \emph {et~al.}(2014)\citenamefont {Fang},
  \citenamefont {Vieira}, \citenamefont {Amorim}, \citenamefont {Mori},\ and\
  \citenamefont {Muggli}}]{fang2014effect}%
  \BibitemOpen
  \bibfield  {author} {\bibinfo {author} {\bibfnamefont {Y.}~\bibnamefont
  {Fang}}, \bibinfo {author} {\bibfnamefont {J.}~\bibnamefont {Vieira}},
  \bibinfo {author} {\bibfnamefont {L.}~\bibnamefont {Amorim}}, \bibinfo
  {author} {\bibfnamefont {W.}~\bibnamefont {Mori}}, \ and\ \bibinfo {author}
  {\bibfnamefont {P.}~\bibnamefont {Muggli}},\ }\href@noop {} {\bibfield
  {journal} {\bibinfo  {journal} {Physics of Plasmas}\ }\textbf {\bibinfo
  {volume} {21}},\ \bibinfo {pages} {056703} (\bibinfo {year}
  {2014})}\BibitemShut {NoStop}%
\bibitem [{\citenamefont {Ferrario}\ \emph {et~al.}(2013)\citenamefont
  {Ferrario}, \citenamefont {Alesini}, \citenamefont {Anania}, \citenamefont
  {Bacci}, \citenamefont {Bellaveglia}, \citenamefont {Bogdanov}, \citenamefont
  {Boni}, \citenamefont {Castellano}, \citenamefont {Chiadroni}, \citenamefont
  {Cianchi} \emph {et~al.}}]{ferrario2013sparc_lab}%
  \BibitemOpen
  \bibfield  {author} {\bibinfo {author} {\bibfnamefont {M.}~\bibnamefont
  {Ferrario}}, \bibinfo {author} {\bibfnamefont {D.}~\bibnamefont {Alesini}},
  \bibinfo {author} {\bibfnamefont {M.}~\bibnamefont {Anania}}, \bibinfo
  {author} {\bibfnamefont {A.}~\bibnamefont {Bacci}}, \bibinfo {author}
  {\bibfnamefont {M.}~\bibnamefont {Bellaveglia}}, \bibinfo {author}
  {\bibfnamefont {O.}~\bibnamefont {Bogdanov}}, \bibinfo {author}
  {\bibfnamefont {R.}~\bibnamefont {Boni}}, \bibinfo {author} {\bibfnamefont
  {M.}~\bibnamefont {Castellano}}, \bibinfo {author} {\bibfnamefont
  {E.}~\bibnamefont {Chiadroni}}, \bibinfo {author} {\bibfnamefont
  {A.}~\bibnamefont {Cianchi}},  \emph {et~al.},\ }\href {\doibase
  10.1016/j.nimb.2013.03.049} {\bibfield  {journal} {\bibinfo  {journal}
  {Nuclear Instruments and Methods B}\ }\textbf {\bibinfo {volume} {309}},\
  \bibinfo {pages} {183} (\bibinfo {year} {2013})}\BibitemShut {NoStop}%
\bibitem [{\citenamefont {Pompili}\ \emph
  {et~al.}(2018{\natexlab{a}})\citenamefont {Pompili}, \citenamefont {Anania},
  \citenamefont {Bellaveglia}, \citenamefont {Biagioni}, \citenamefont {Bini},
  \citenamefont {Bisesto}, \citenamefont {Chiadroni}, \citenamefont {Cianchi},
  \citenamefont {Costa}, \citenamefont {Giovenale}, \citenamefont {Ferrario},
  \citenamefont {Filippi}, \citenamefont {Gallo}, \citenamefont {Giribono},
  \citenamefont {Lollo}, \citenamefont {Marocchino}, \citenamefont
  {Martinelli}, \citenamefont {Mostacci}, \citenamefont {Pirro}, \citenamefont
  {Romeo}, \citenamefont {Scifo}, \citenamefont {Shpakov}, \citenamefont
  {Vaccarezza}, \citenamefont {Villa},\ and\ \citenamefont
  {Zigler}}]{pompili2018recentres}%
  \BibitemOpen
  \bibfield  {author} {\bibinfo {author} {\bibfnamefont {R.}~\bibnamefont
  {Pompili}}, \bibinfo {author} {\bibfnamefont {M.}~\bibnamefont {Anania}},
  \bibinfo {author} {\bibfnamefont {M.}~\bibnamefont {Bellaveglia}}, \bibinfo
  {author} {\bibfnamefont {A.}~\bibnamefont {Biagioni}}, \bibinfo {author}
  {\bibfnamefont {S.}~\bibnamefont {Bini}}, \bibinfo {author} {\bibfnamefont
  {F.}~\bibnamefont {Bisesto}}, \bibinfo {author} {\bibfnamefont
  {E.}~\bibnamefont {Chiadroni}}, \bibinfo {author} {\bibfnamefont
  {A.}~\bibnamefont {Cianchi}}, \bibinfo {author} {\bibfnamefont
  {G.}~\bibnamefont {Costa}}, \bibinfo {author} {\bibfnamefont {D.~D.}\
  \bibnamefont {Giovenale}}, \bibinfo {author} {\bibfnamefont {M.}~\bibnamefont
  {Ferrario}}, \bibinfo {author} {\bibfnamefont {F.}~\bibnamefont {Filippi}},
  \bibinfo {author} {\bibfnamefont {A.}~\bibnamefont {Gallo}}, \bibinfo
  {author} {\bibfnamefont {A.}~\bibnamefont {Giribono}}, \bibinfo {author}
  {\bibfnamefont {V.}~\bibnamefont {Lollo}}, \bibinfo {author} {\bibfnamefont
  {A.}~\bibnamefont {Marocchino}}, \bibinfo {author} {\bibfnamefont
  {V.}~\bibnamefont {Martinelli}}, \bibinfo {author} {\bibfnamefont
  {A.}~\bibnamefont {Mostacci}}, \bibinfo {author} {\bibfnamefont {G.~D.}\
  \bibnamefont {Pirro}}, \bibinfo {author} {\bibfnamefont {S.}~\bibnamefont
  {Romeo}}, \bibinfo {author} {\bibfnamefont {J.}~\bibnamefont {Scifo}},
  \bibinfo {author} {\bibfnamefont {V.}~\bibnamefont {Shpakov}}, \bibinfo
  {author} {\bibfnamefont {C.}~\bibnamefont {Vaccarezza}}, \bibinfo {author}
  {\bibfnamefont {F.}~\bibnamefont {Villa}}, \ and\ \bibinfo {author}
  {\bibfnamefont {A.}~\bibnamefont {Zigler}},\ }\href {\doibase
  https://doi.org/10.1016/j.nima.2018.01.071} {\bibfield  {journal} {\bibinfo
  {journal} {Nuclear Instruments and Methods in Physics Research Section A:
  Accelerators, Spectrometers, Detectors and Associated Equipment}\ } (\bibinfo
  {year} {2018}{\natexlab{a}}),\
  https://doi.org/10.1016/j.nima.2018.01.071}\BibitemShut {NoStop}%
\bibitem [{\citenamefont {Pompili}\ \emph
  {et~al.}(2018{\natexlab{b}})\citenamefont {Pompili}, \citenamefont {Anania},
  \citenamefont {Bellaveglia}, \citenamefont {Biagioni}, \citenamefont {Bini},
  \citenamefont {Bisesto}, \citenamefont {Brentegani}, \citenamefont
  {Cardelli}, \citenamefont {Castorina}, \citenamefont {Chiadroni},
  \citenamefont {Cianchi}, \citenamefont {Coiro}, \citenamefont {Costa},
  \citenamefont {Croia}, \citenamefont {Di~Giovenale}, \citenamefont
  {Ferrario}, \citenamefont {Filippi}, \citenamefont {Giribono}, \citenamefont
  {Lollo}, \citenamefont {Marocchino}, \citenamefont {Marongiu}, \citenamefont
  {Martinelli}, \citenamefont {Mostacci}, \citenamefont {Pellegrini},
  \citenamefont {Piersanti}, \citenamefont {Di~Pirro}, \citenamefont {Romeo},
  \citenamefont {Rossi}, \citenamefont {Scifo}, \citenamefont {Shpakov},
  \citenamefont {Stella}, \citenamefont {Vaccarezza}, \citenamefont {Villa},\
  and\ \citenamefont {Zigler}}]{PhysRevLett.121.174801}%
  \BibitemOpen
  \bibfield  {author} {\bibinfo {author} {\bibfnamefont {R.}~\bibnamefont
  {Pompili}}, \bibinfo {author} {\bibfnamefont {M.~P.}\ \bibnamefont {Anania}},
  \bibinfo {author} {\bibfnamefont {M.}~\bibnamefont {Bellaveglia}}, \bibinfo
  {author} {\bibfnamefont {A.}~\bibnamefont {Biagioni}}, \bibinfo {author}
  {\bibfnamefont {S.}~\bibnamefont {Bini}}, \bibinfo {author} {\bibfnamefont
  {F.}~\bibnamefont {Bisesto}}, \bibinfo {author} {\bibfnamefont
  {E.}~\bibnamefont {Brentegani}}, \bibinfo {author} {\bibfnamefont
  {F.}~\bibnamefont {Cardelli}}, \bibinfo {author} {\bibfnamefont
  {G.}~\bibnamefont {Castorina}}, \bibinfo {author} {\bibfnamefont
  {E.}~\bibnamefont {Chiadroni}}, \bibinfo {author} {\bibfnamefont
  {A.}~\bibnamefont {Cianchi}}, \bibinfo {author} {\bibfnamefont
  {O.}~\bibnamefont {Coiro}}, \bibinfo {author} {\bibfnamefont
  {G.}~\bibnamefont {Costa}}, \bibinfo {author} {\bibfnamefont
  {M.}~\bibnamefont {Croia}}, \bibinfo {author} {\bibfnamefont
  {D.}~\bibnamefont {Di~Giovenale}}, \bibinfo {author} {\bibfnamefont
  {M.}~\bibnamefont {Ferrario}}, \bibinfo {author} {\bibfnamefont
  {F.}~\bibnamefont {Filippi}}, \bibinfo {author} {\bibfnamefont
  {A.}~\bibnamefont {Giribono}}, \bibinfo {author} {\bibfnamefont
  {V.}~\bibnamefont {Lollo}}, \bibinfo {author} {\bibfnamefont
  {A.}~\bibnamefont {Marocchino}}, \bibinfo {author} {\bibfnamefont
  {M.}~\bibnamefont {Marongiu}}, \bibinfo {author} {\bibfnamefont
  {V.}~\bibnamefont {Martinelli}}, \bibinfo {author} {\bibfnamefont
  {A.}~\bibnamefont {Mostacci}}, \bibinfo {author} {\bibfnamefont
  {D.}~\bibnamefont {Pellegrini}}, \bibinfo {author} {\bibfnamefont
  {L.}~\bibnamefont {Piersanti}}, \bibinfo {author} {\bibfnamefont
  {G.}~\bibnamefont {Di~Pirro}}, \bibinfo {author} {\bibfnamefont
  {S.}~\bibnamefont {Romeo}}, \bibinfo {author} {\bibfnamefont {A.~R.}\
  \bibnamefont {Rossi}}, \bibinfo {author} {\bibfnamefont {J.}~\bibnamefont
  {Scifo}}, \bibinfo {author} {\bibfnamefont {V.}~\bibnamefont {Shpakov}},
  \bibinfo {author} {\bibfnamefont {A.}~\bibnamefont {Stella}}, \bibinfo
  {author} {\bibfnamefont {C.}~\bibnamefont {Vaccarezza}}, \bibinfo {author}
  {\bibfnamefont {F.}~\bibnamefont {Villa}}, \ and\ \bibinfo {author}
  {\bibfnamefont {A.}~\bibnamefont {Zigler}},\ }\href {\doibase
  10.1103/PhysRevLett.121.174801} {\bibfield  {journal} {\bibinfo  {journal}
  {Phys. Rev. Lett.}\ }\textbf {\bibinfo {volume} {121}},\ \bibinfo {pages}
  {174801} (\bibinfo {year} {2018}{\natexlab{b}})}\BibitemShut {NoStop}%
\bibitem [{\citenamefont {Marocchino}\ \emph {et~al.}(2017)\citenamefont
  {Marocchino}, \citenamefont {Anania}, \citenamefont {Bellaveglia},
  \citenamefont {Biagioni}, \citenamefont {Bini}, \citenamefont {Bisesto},
  \citenamefont {Brentegani}, \citenamefont {Chiadroni}, \citenamefont
  {Cianchi}, \citenamefont {Croia}, \citenamefont {Giovenale}, \citenamefont
  {Ferrario}, \citenamefont {Filippi}, \citenamefont {Giribono}, \citenamefont
  {Lollo}, \citenamefont {Marongiu}, \citenamefont {Mostacci}, \citenamefont
  {Pirro}, \citenamefont {Pompili}, \citenamefont {Romeo}, \citenamefont
  {Rossi}, \citenamefont {Scifo}, \citenamefont {Shpakov}, \citenamefont
  {Vaccarezza}, \citenamefont {Villa},\ and\ \citenamefont
  {Zigler}}]{lens_alberto}%
  \BibitemOpen
  \bibfield  {author} {\bibinfo {author} {\bibfnamefont {A.}~\bibnamefont
  {Marocchino}}, \bibinfo {author} {\bibfnamefont {M.~P.}\ \bibnamefont
  {Anania}}, \bibinfo {author} {\bibfnamefont {M.}~\bibnamefont {Bellaveglia}},
  \bibinfo {author} {\bibfnamefont {A.}~\bibnamefont {Biagioni}}, \bibinfo
  {author} {\bibfnamefont {S.}~\bibnamefont {Bini}}, \bibinfo {author}
  {\bibfnamefont {F.}~\bibnamefont {Bisesto}}, \bibinfo {author} {\bibfnamefont
  {E.}~\bibnamefont {Brentegani}}, \bibinfo {author} {\bibfnamefont
  {E.}~\bibnamefont {Chiadroni}}, \bibinfo {author} {\bibfnamefont
  {A.}~\bibnamefont {Cianchi}}, \bibinfo {author} {\bibfnamefont
  {M.}~\bibnamefont {Croia}}, \bibinfo {author} {\bibfnamefont {D.~D.}\
  \bibnamefont {Giovenale}}, \bibinfo {author} {\bibfnamefont {M.}~\bibnamefont
  {Ferrario}}, \bibinfo {author} {\bibfnamefont {F.}~\bibnamefont {Filippi}},
  \bibinfo {author} {\bibfnamefont {A.}~\bibnamefont {Giribono}}, \bibinfo
  {author} {\bibfnamefont {V.}~\bibnamefont {Lollo}}, \bibinfo {author}
  {\bibfnamefont {M.}~\bibnamefont {Marongiu}}, \bibinfo {author}
  {\bibfnamefont {A.}~\bibnamefont {Mostacci}}, \bibinfo {author}
  {\bibfnamefont {G.~D.}\ \bibnamefont {Pirro}}, \bibinfo {author}
  {\bibfnamefont {R.}~\bibnamefont {Pompili}}, \bibinfo {author} {\bibfnamefont
  {S.}~\bibnamefont {Romeo}}, \bibinfo {author} {\bibfnamefont {A.~R.}\
  \bibnamefont {Rossi}}, \bibinfo {author} {\bibfnamefont {J.}~\bibnamefont
  {Scifo}}, \bibinfo {author} {\bibfnamefont {V.}~\bibnamefont {Shpakov}},
  \bibinfo {author} {\bibfnamefont {C.}~\bibnamefont {Vaccarezza}}, \bibinfo
  {author} {\bibfnamefont {F.}~\bibnamefont {Villa}}, \ and\ \bibinfo {author}
  {\bibfnamefont {A.}~\bibnamefont {Zigler}},\ }\href@noop {} {\bibfield
  {journal} {\bibinfo  {journal} {Applied Physics Letters}\ }\textbf {\bibinfo
  {volume} {111}},\ \bibinfo {pages} {184101} (\bibinfo {year}
  {2017})}\BibitemShut {NoStop}%
\bibitem [{\citenamefont {Pompili}\ \emph {et~al.}(2017)\citenamefont
  {Pompili}, \citenamefont {Anania}, \citenamefont {Bellaveglia}, \citenamefont
  {Biagioni}, \citenamefont {Bini}, \citenamefont {Bisesto}, \citenamefont
  {Brentegani}, \citenamefont {Castorina}, \citenamefont {Chiadroni},
  \citenamefont {Cianchi} \emph {et~al.}}]{pompili2017experimental}%
  \BibitemOpen
  \bibfield  {author} {\bibinfo {author} {\bibfnamefont {R.}~\bibnamefont
  {Pompili}}, \bibinfo {author} {\bibfnamefont {M.}~\bibnamefont {Anania}},
  \bibinfo {author} {\bibfnamefont {M.}~\bibnamefont {Bellaveglia}}, \bibinfo
  {author} {\bibfnamefont {A.}~\bibnamefont {Biagioni}}, \bibinfo {author}
  {\bibfnamefont {S.}~\bibnamefont {Bini}}, \bibinfo {author} {\bibfnamefont
  {F.}~\bibnamefont {Bisesto}}, \bibinfo {author} {\bibfnamefont
  {E.}~\bibnamefont {Brentegani}}, \bibinfo {author} {\bibfnamefont
  {G.}~\bibnamefont {Castorina}}, \bibinfo {author} {\bibfnamefont
  {E.}~\bibnamefont {Chiadroni}}, \bibinfo {author} {\bibfnamefont
  {A.}~\bibnamefont {Cianchi}},  \emph {et~al.},\ }\href@noop {} {\bibfield
  {journal} {\bibinfo  {journal} {Applied Physics Letters}\ }\textbf {\bibinfo
  {volume} {110}},\ \bibinfo {pages} {104101} (\bibinfo {year}
  {2017})}\BibitemShut {NoStop}%
\bibitem [{\citenamefont {Alesini}\ \emph {et~al.}(2003)\citenamefont
  {Alesini}, \citenamefont {Bertolucci}, \citenamefont {Biagini}, \citenamefont
  {Biscari}, \citenamefont {Boni}, \citenamefont {Boscolo}, \citenamefont
  {Castellano}, \citenamefont {Clozza}, \citenamefont {Di~Pirro}, \citenamefont
  {Drago} \emph {et~al.}}]{Alesini2003345}%
  \BibitemOpen
  \bibfield  {author} {\bibinfo {author} {\bibfnamefont {D.}~\bibnamefont
  {Alesini}}, \bibinfo {author} {\bibfnamefont {S.}~\bibnamefont {Bertolucci}},
  \bibinfo {author} {\bibfnamefont {M.}~\bibnamefont {Biagini}}, \bibinfo
  {author} {\bibfnamefont {C.}~\bibnamefont {Biscari}}, \bibinfo {author}
  {\bibfnamefont {R.}~\bibnamefont {Boni}}, \bibinfo {author} {\bibfnamefont
  {M.}~\bibnamefont {Boscolo}}, \bibinfo {author} {\bibfnamefont
  {M.}~\bibnamefont {Castellano}}, \bibinfo {author} {\bibfnamefont
  {A.}~\bibnamefont {Clozza}}, \bibinfo {author} {\bibfnamefont
  {G.}~\bibnamefont {Di~Pirro}}, \bibinfo {author} {\bibfnamefont
  {A.}~\bibnamefont {Drago}},  \emph {et~al.},\ }\href {\doibase
  http://dx.doi.org/10.1016/S0168-9002(03)00943-4} {\bibfield  {journal}
  {\bibinfo  {journal} {Nuclear Instruments and Methods A}\ }\textbf {\bibinfo
  {volume} {507}},\ \bibinfo {pages} {345 } (\bibinfo {year}
  {2003})}\BibitemShut {NoStop}%
\bibitem [{\citenamefont {Chiadroni}\ \emph
  {et~al.}(2013{\natexlab{b}})\citenamefont {Chiadroni}, \citenamefont {Bacci},
  \citenamefont {Bellaveglia}, \citenamefont {Boscolo}, \citenamefont
  {Castellano}, \citenamefont {Cultrera}, \citenamefont {Di~Pirro},
  \citenamefont {Ferrario}, \citenamefont {Ficcadenti}, \citenamefont
  {Filippetto} \emph {et~al.}}]{chiadroni2013sparc}%
  \BibitemOpen
  \bibfield  {author} {\bibinfo {author} {\bibfnamefont {E.}~\bibnamefont
  {Chiadroni}}, \bibinfo {author} {\bibfnamefont {A.}~\bibnamefont {Bacci}},
  \bibinfo {author} {\bibfnamefont {M.}~\bibnamefont {Bellaveglia}}, \bibinfo
  {author} {\bibfnamefont {M.}~\bibnamefont {Boscolo}}, \bibinfo {author}
  {\bibfnamefont {M.}~\bibnamefont {Castellano}}, \bibinfo {author}
  {\bibfnamefont {L.}~\bibnamefont {Cultrera}}, \bibinfo {author}
  {\bibfnamefont {G.}~\bibnamefont {Di~Pirro}}, \bibinfo {author}
  {\bibfnamefont {M.}~\bibnamefont {Ferrario}}, \bibinfo {author}
  {\bibfnamefont {L.}~\bibnamefont {Ficcadenti}}, \bibinfo {author}
  {\bibfnamefont {D.}~\bibnamefont {Filippetto}},  \emph {et~al.},\ }\href@noop
  {} {\bibfield  {journal} {\bibinfo  {journal} {Applied Physics Letters}\
  }\textbf {\bibinfo {volume} {102}},\ \bibinfo {pages} {094101} (\bibinfo
  {year} {2013}{\natexlab{b}})}\BibitemShut {NoStop}%
\bibitem [{\citenamefont {Cianchi}\ \emph {et~al.}(2008)\citenamefont
  {Cianchi}, \citenamefont {Alesini}, \citenamefont {Bacci}, \citenamefont
  {Bellaveglia}, \citenamefont {Boni}, \citenamefont {Boscolo}, \citenamefont
  {Castellano}, \citenamefont {Catani}, \citenamefont {Chiadroni},
  \citenamefont {Cialdi} \emph {et~al.}}]{cianchi2008high}%
  \BibitemOpen
  \bibfield  {author} {\bibinfo {author} {\bibfnamefont {A.}~\bibnamefont
  {Cianchi}}, \bibinfo {author} {\bibfnamefont {D.}~\bibnamefont {Alesini}},
  \bibinfo {author} {\bibfnamefont {A.}~\bibnamefont {Bacci}}, \bibinfo
  {author} {\bibfnamefont {M.}~\bibnamefont {Bellaveglia}}, \bibinfo {author}
  {\bibfnamefont {R.}~\bibnamefont {Boni}}, \bibinfo {author} {\bibfnamefont
  {M.}~\bibnamefont {Boscolo}}, \bibinfo {author} {\bibfnamefont
  {M.}~\bibnamefont {Castellano}}, \bibinfo {author} {\bibfnamefont
  {L.}~\bibnamefont {Catani}}, \bibinfo {author} {\bibfnamefont
  {E.}~\bibnamefont {Chiadroni}}, \bibinfo {author} {\bibfnamefont
  {S.}~\bibnamefont {Cialdi}},  \emph {et~al.},\ }\href@noop {} {\bibfield
  {journal} {\bibinfo  {journal} {Physical Review Special Topics-Accelerators
  and Beams}\ }\textbf {\bibinfo {volume} {11}},\ \bibinfo {pages} {032801}
  (\bibinfo {year} {2008})}\BibitemShut {NoStop}%
\bibitem [{\citenamefont {Ferrario}\ \emph {et~al.}(2010)\citenamefont
  {Ferrario}, \citenamefont {Alesini}, \citenamefont {Bacci}, \citenamefont
  {Bellaveglia}, \citenamefont {Boni}, \citenamefont {Boscolo}, \citenamefont
  {Castellano}, \citenamefont {Chiadroni}, \citenamefont {Cianchi},
  \citenamefont {Cultrera} \emph {et~al.}}]{ferrario2010experimental}%
  \BibitemOpen
  \bibfield  {author} {\bibinfo {author} {\bibfnamefont {M.}~\bibnamefont
  {Ferrario}}, \bibinfo {author} {\bibfnamefont {D.}~\bibnamefont {Alesini}},
  \bibinfo {author} {\bibfnamefont {A.}~\bibnamefont {Bacci}}, \bibinfo
  {author} {\bibfnamefont {M.}~\bibnamefont {Bellaveglia}}, \bibinfo {author}
  {\bibfnamefont {R.}~\bibnamefont {Boni}}, \bibinfo {author} {\bibfnamefont
  {M.}~\bibnamefont {Boscolo}}, \bibinfo {author} {\bibfnamefont
  {M.}~\bibnamefont {Castellano}}, \bibinfo {author} {\bibfnamefont
  {E.}~\bibnamefont {Chiadroni}}, \bibinfo {author} {\bibfnamefont
  {A.}~\bibnamefont {Cianchi}}, \bibinfo {author} {\bibfnamefont
  {L.}~\bibnamefont {Cultrera}},  \emph {et~al.},\ }\href@noop {} {\bibfield
  {journal} {\bibinfo  {journal} {Physical review letters}\ }\textbf {\bibinfo
  {volume} {104}},\ \bibinfo {pages} {054801} (\bibinfo {year}
  {2010})}\BibitemShut {NoStop}%
\bibitem [{\citenamefont {Pompili}\ \emph
  {et~al.}(2016{\natexlab{a}})\citenamefont {Pompili}, \citenamefont {Anania},
  \citenamefont {Bellaveglia}, \citenamefont {Biagioni}, \citenamefont
  {Bisesto}, \citenamefont {Chiadroni}, \citenamefont {Cianchi}, \citenamefont
  {Croia}, \citenamefont {Curcio}, \citenamefont {Di~Giovenale} \emph
  {et~al.}}]{pompili2016beam}%
  \BibitemOpen
  \bibfield  {author} {\bibinfo {author} {\bibfnamefont {R.}~\bibnamefont
  {Pompili}}, \bibinfo {author} {\bibfnamefont {M.}~\bibnamefont {Anania}},
  \bibinfo {author} {\bibfnamefont {M.}~\bibnamefont {Bellaveglia}}, \bibinfo
  {author} {\bibfnamefont {A.}~\bibnamefont {Biagioni}}, \bibinfo {author}
  {\bibfnamefont {F.}~\bibnamefont {Bisesto}}, \bibinfo {author} {\bibfnamefont
  {E.}~\bibnamefont {Chiadroni}}, \bibinfo {author} {\bibfnamefont
  {A.}~\bibnamefont {Cianchi}}, \bibinfo {author} {\bibfnamefont
  {M.}~\bibnamefont {Croia}}, \bibinfo {author} {\bibfnamefont
  {A.}~\bibnamefont {Curcio}}, \bibinfo {author} {\bibfnamefont
  {D.}~\bibnamefont {Di~Giovenale}},  \emph {et~al.},\ }\href@noop {}
  {\bibfield  {journal} {\bibinfo  {journal} {Nuclear Instruments and Methods
  in Physics Research Section A: Accelerators, Spectrometers, Detectors and
  Associated Equipment}\ }\textbf {\bibinfo {volume} {829}},\ \bibinfo {pages}
  {17} (\bibinfo {year} {2016}{\natexlab{a}})}\BibitemShut {NoStop}%
\bibitem [{\citenamefont {Anania}\ \emph {et~al.}(2016)\citenamefont {Anania},
  \citenamefont {Biagioni}, \citenamefont {Chiadroni}, \citenamefont {Cianchi},
  \citenamefont {Croia}, \citenamefont {Curcio}, \citenamefont {Di~Giovenale},
  \citenamefont {Di~Pirro}, \citenamefont {Filippi}, \citenamefont {Ghigo}
  \emph {et~al.}}]{anania2016plasma}%
  \BibitemOpen
  \bibfield  {author} {\bibinfo {author} {\bibfnamefont {M.}~\bibnamefont
  {Anania}}, \bibinfo {author} {\bibfnamefont {A.}~\bibnamefont {Biagioni}},
  \bibinfo {author} {\bibfnamefont {E.}~\bibnamefont {Chiadroni}}, \bibinfo
  {author} {\bibfnamefont {A.}~\bibnamefont {Cianchi}}, \bibinfo {author}
  {\bibfnamefont {M.}~\bibnamefont {Croia}}, \bibinfo {author} {\bibfnamefont
  {A.}~\bibnamefont {Curcio}}, \bibinfo {author} {\bibfnamefont
  {D.}~\bibnamefont {Di~Giovenale}}, \bibinfo {author} {\bibfnamefont
  {G.}~\bibnamefont {Di~Pirro}}, \bibinfo {author} {\bibfnamefont
  {F.}~\bibnamefont {Filippi}}, \bibinfo {author} {\bibfnamefont
  {A.}~\bibnamefont {Ghigo}},  \emph {et~al.},\ }\href@noop {} {\bibfield
  {journal} {\bibinfo  {journal} {Nuclear Instruments and Methods in Physics
  Research Section A: Accelerators, Spectrometers, Detectors and Associated
  Equipment}\ } (\bibinfo {year} {2016})}\BibitemShut {NoStop}%
\bibitem [{\citenamefont {Filippi}\ \emph
  {et~al.}(2016{\natexlab{a}})\citenamefont {Filippi}, \citenamefont {Anania},
  \citenamefont {Biagioni}, \citenamefont {Chiadroni}, \citenamefont {Cianchi},
  \citenamefont {Ferrario}, \citenamefont {Mostacci}, \citenamefont {Palumbo},\
  and\ \citenamefont {Zigler}}]{filippi2016spectroscopic}%
  \BibitemOpen
  \bibfield  {author} {\bibinfo {author} {\bibfnamefont {F.}~\bibnamefont
  {Filippi}}, \bibinfo {author} {\bibfnamefont {M.}~\bibnamefont {Anania}},
  \bibinfo {author} {\bibfnamefont {A.}~\bibnamefont {Biagioni}}, \bibinfo
  {author} {\bibfnamefont {E.}~\bibnamefont {Chiadroni}}, \bibinfo {author}
  {\bibfnamefont {A.}~\bibnamefont {Cianchi}}, \bibinfo {author} {\bibfnamefont
  {M.}~\bibnamefont {Ferrario}}, \bibinfo {author} {\bibfnamefont
  {A.}~\bibnamefont {Mostacci}}, \bibinfo {author} {\bibfnamefont
  {L.}~\bibnamefont {Palumbo}}, \ and\ \bibinfo {author} {\bibfnamefont
  {A.}~\bibnamefont {Zigler}},\ }\href@noop {} {\bibfield  {journal} {\bibinfo
  {journal} {Journal of Instrumentation}\ }\textbf {\bibinfo {volume} {11}},\
  \bibinfo {pages} {C09015} (\bibinfo {year} {2016}{\natexlab{a}})}\BibitemShut
  {NoStop}%
\bibitem [{\citenamefont {Serafini}\ and\ \citenamefont
  {Ferrario}(2001)}]{serafini2001velocity}%
  \BibitemOpen
  \bibfield  {author} {\bibinfo {author} {\bibfnamefont {L.}~\bibnamefont
  {Serafini}}\ and\ \bibinfo {author} {\bibfnamefont {M.}~\bibnamefont
  {Ferrario}},\ }in\ \href@noop {} {\emph {\bibinfo {booktitle} {American
  Institute of Physics Conference Series}}},\ Vol.\ \bibinfo {volume} {581}\
  (\bibinfo {year} {2001})\ pp.\ \bibinfo {pages} {87--106}\BibitemShut
  {NoStop}%
\bibitem [{\citenamefont {Pompili}\ \emph
  {et~al.}(2016{\natexlab{b}})\citenamefont {Pompili}, \citenamefont {Anania},
  \citenamefont {Bellaveglia}, \citenamefont {Biagioni}, \citenamefont
  {Castorina}, \citenamefont {Chiadroni}, \citenamefont {Cianchi},
  \citenamefont {Croia}, \citenamefont {Giovenale}, \citenamefont {Ferrario},
  \citenamefont {Filippi}, \citenamefont {Gallo}, \citenamefont {Gatti},
  \citenamefont {Giorgianni}, \citenamefont {Giribono}, \citenamefont {Li},
  \citenamefont {Lupi}, \citenamefont {Mostacci}, \citenamefont {Petrarca},
  \citenamefont {Piersanti}, \citenamefont {Pirro}, \citenamefont {Romeo},
  \citenamefont {Scifo}, \citenamefont {Shpakov}, \citenamefont {Vaccarezza},\
  and\ \citenamefont {Villa}}]{eos_jitter}%
  \BibitemOpen
  \bibfield  {author} {\bibinfo {author} {\bibfnamefont {R.}~\bibnamefont
  {Pompili}}, \bibinfo {author} {\bibfnamefont {M.~P.}\ \bibnamefont {Anania}},
  \bibinfo {author} {\bibfnamefont {M.}~\bibnamefont {Bellaveglia}}, \bibinfo
  {author} {\bibfnamefont {A.}~\bibnamefont {Biagioni}}, \bibinfo {author}
  {\bibfnamefont {G.}~\bibnamefont {Castorina}}, \bibinfo {author}
  {\bibfnamefont {E.}~\bibnamefont {Chiadroni}}, \bibinfo {author}
  {\bibfnamefont {A.}~\bibnamefont {Cianchi}}, \bibinfo {author} {\bibfnamefont
  {M.}~\bibnamefont {Croia}}, \bibinfo {author} {\bibfnamefont {D.~D.}\
  \bibnamefont {Giovenale}}, \bibinfo {author} {\bibfnamefont {M.}~\bibnamefont
  {Ferrario}}, \bibinfo {author} {\bibfnamefont {F.}~\bibnamefont {Filippi}},
  \bibinfo {author} {\bibfnamefont {A.}~\bibnamefont {Gallo}}, \bibinfo
  {author} {\bibfnamefont {G.}~\bibnamefont {Gatti}}, \bibinfo {author}
  {\bibfnamefont {F.}~\bibnamefont {Giorgianni}}, \bibinfo {author}
  {\bibfnamefont {A.}~\bibnamefont {Giribono}}, \bibinfo {author}
  {\bibfnamefont {W.}~\bibnamefont {Li}}, \bibinfo {author} {\bibfnamefont
  {S.}~\bibnamefont {Lupi}}, \bibinfo {author} {\bibfnamefont {A.}~\bibnamefont
  {Mostacci}}, \bibinfo {author} {\bibfnamefont {M.}~\bibnamefont {Petrarca}},
  \bibinfo {author} {\bibfnamefont {L.}~\bibnamefont {Piersanti}}, \bibinfo
  {author} {\bibfnamefont {G.~D.}\ \bibnamefont {Pirro}}, \bibinfo {author}
  {\bibfnamefont {S.}~\bibnamefont {Romeo}}, \bibinfo {author} {\bibfnamefont
  {J.}~\bibnamefont {Scifo}}, \bibinfo {author} {\bibfnamefont
  {V.}~\bibnamefont {Shpakov}}, \bibinfo {author} {\bibfnamefont
  {C.}~\bibnamefont {Vaccarezza}}, \ and\ \bibinfo {author} {\bibfnamefont
  {F.}~\bibnamefont {Villa}},\ }\href
  {http://stacks.iop.org/1367-2630/18/i=8/a=083033} {\bibfield  {journal}
  {\bibinfo  {journal} {New Journal of Physics}\ }\textbf {\bibinfo {volume}
  {18}},\ \bibinfo {pages} {083033} (\bibinfo {year}
  {2016}{\natexlab{b}})}\BibitemShut {NoStop}%
\bibitem [{\citenamefont {Alesini}\ \emph {et~al.}(2006)\citenamefont
  {Alesini}, \citenamefont {Di~Pirro}, \citenamefont {Ficcadenti},
  \citenamefont {Mostacci}, \citenamefont {Palumbo}, \citenamefont
  {Rosenzweig},\ and\ \citenamefont {Vaccarezza}}]{alesini2006rf}%
  \BibitemOpen
  \bibfield  {author} {\bibinfo {author} {\bibfnamefont {D.}~\bibnamefont
  {Alesini}}, \bibinfo {author} {\bibfnamefont {G.}~\bibnamefont {Di~Pirro}},
  \bibinfo {author} {\bibfnamefont {L.}~\bibnamefont {Ficcadenti}}, \bibinfo
  {author} {\bibfnamefont {A.}~\bibnamefont {Mostacci}}, \bibinfo {author}
  {\bibfnamefont {L.}~\bibnamefont {Palumbo}}, \bibinfo {author} {\bibfnamefont
  {J.}~\bibnamefont {Rosenzweig}}, \ and\ \bibinfo {author} {\bibfnamefont
  {C.}~\bibnamefont {Vaccarezza}},\ }\href@noop {} {\bibfield  {journal}
  {\bibinfo  {journal} {Nuclear Instruments and Methods in Physics Research
  Section A: Accelerators, Spectrometers, Detectors and Associated Equipment}\
  }\textbf {\bibinfo {volume} {568}},\ \bibinfo {pages} {488} (\bibinfo {year}
  {2006})}\BibitemShut {NoStop}%
\bibitem [{\citenamefont {Pompili}\ \emph
  {et~al.}(2018{\natexlab{c}})\citenamefont {Pompili}, \citenamefont {Anania},
  \citenamefont {Chiadroni}, \citenamefont {Cianchi}, \citenamefont {Ferrario},
  \citenamefont {Lollo}, \citenamefont {Notargiacomo}, \citenamefont {Picardi},
  \citenamefont {Ronsivalle}, \citenamefont {Rosenzweig} \emph
  {et~al.}}]{pompili2018compact}%
  \BibitemOpen
  \bibfield  {author} {\bibinfo {author} {\bibfnamefont {R.}~\bibnamefont
  {Pompili}}, \bibinfo {author} {\bibfnamefont {M.}~\bibnamefont {Anania}},
  \bibinfo {author} {\bibfnamefont {E.}~\bibnamefont {Chiadroni}}, \bibinfo
  {author} {\bibfnamefont {A.}~\bibnamefont {Cianchi}}, \bibinfo {author}
  {\bibfnamefont {M.}~\bibnamefont {Ferrario}}, \bibinfo {author}
  {\bibfnamefont {V.}~\bibnamefont {Lollo}}, \bibinfo {author} {\bibfnamefont
  {A.}~\bibnamefont {Notargiacomo}}, \bibinfo {author} {\bibfnamefont
  {L.}~\bibnamefont {Picardi}}, \bibinfo {author} {\bibfnamefont
  {C.}~\bibnamefont {Ronsivalle}}, \bibinfo {author} {\bibfnamefont
  {J.}~\bibnamefont {Rosenzweig}},  \emph {et~al.},\ }\href@noop {} {\bibfield
  {journal} {\bibinfo  {journal} {Review of Scientific Instruments}\ }\textbf
  {\bibinfo {volume} {89}},\ \bibinfo {pages} {033302} (\bibinfo {year}
  {2018}{\natexlab{c}})}\BibitemShut {NoStop}%
\bibitem [{\citenamefont {Filippi}\ \emph
  {et~al.}(2016{\natexlab{b}})\citenamefont {Filippi}, \citenamefont {Anania},
  \citenamefont {Bellaveglia}, \citenamefont {Biagioni}, \citenamefont
  {Chiadroni}, \citenamefont {Cianchi}, \citenamefont {Di~Giovenale},
  \citenamefont {Di~Pirro}, \citenamefont {Ferrario}, \citenamefont {Mostacci}
  \emph {et~al.}}]{filippi2016plasma}%
  \BibitemOpen
  \bibfield  {author} {\bibinfo {author} {\bibfnamefont {F.}~\bibnamefont
  {Filippi}}, \bibinfo {author} {\bibfnamefont {M.}~\bibnamefont {Anania}},
  \bibinfo {author} {\bibfnamefont {M.}~\bibnamefont {Bellaveglia}}, \bibinfo
  {author} {\bibfnamefont {A.}~\bibnamefont {Biagioni}}, \bibinfo {author}
  {\bibfnamefont {E.}~\bibnamefont {Chiadroni}}, \bibinfo {author}
  {\bibfnamefont {A.}~\bibnamefont {Cianchi}}, \bibinfo {author} {\bibfnamefont
  {D.}~\bibnamefont {Di~Giovenale}}, \bibinfo {author} {\bibfnamefont
  {G.}~\bibnamefont {Di~Pirro}}, \bibinfo {author} {\bibfnamefont
  {M.}~\bibnamefont {Ferrario}}, \bibinfo {author} {\bibfnamefont
  {A.}~\bibnamefont {Mostacci}},  \emph {et~al.},\ }\href@noop {} {\bibfield
  {journal} {\bibinfo  {journal} {Nuclear Instruments and Methods in Physics
  Research Section A: Accelerators, Spectrometers, Detectors and Associated
  Equipment}\ } (\bibinfo {year} {2016}{\natexlab{b}})}\BibitemShut {NoStop}%
\bibitem [{\citenamefont {Johnson}\ and\ \citenamefont
  {Hinnov}(1973)}]{johnson1973ionization}%
  \BibitemOpen
  \bibfield  {author} {\bibinfo {author} {\bibfnamefont {L.}~\bibnamefont
  {Johnson}}\ and\ \bibinfo {author} {\bibfnamefont {E.}~\bibnamefont
  {Hinnov}},\ }\href@noop {} {\bibfield  {journal} {\bibinfo  {journal}
  {Journal of Quantitative Spectroscopy and Radiative Transfer}\ }\textbf
  {\bibinfo {volume} {13}},\ \bibinfo {pages} {333} (\bibinfo {year}
  {1973})}\BibitemShut {NoStop}%
\bibitem [{\citenamefont {Lu}\ \emph {et~al.}(2005)\citenamefont {Lu},
  \citenamefont {Huang}, \citenamefont {Zhou}, \citenamefont {Mori},\ and\
  \citenamefont {Katsouleas}}]{lu2005limits}%
  \BibitemOpen
  \bibfield  {author} {\bibinfo {author} {\bibfnamefont {W.}~\bibnamefont
  {Lu}}, \bibinfo {author} {\bibfnamefont {C.}~\bibnamefont {Huang}}, \bibinfo
  {author} {\bibfnamefont {M.}~\bibnamefont {Zhou}}, \bibinfo {author}
  {\bibfnamefont {W.}~\bibnamefont {Mori}}, \ and\ \bibinfo {author}
  {\bibfnamefont {T.}~\bibnamefont {Katsouleas}},\ }\href@noop {} {\bibfield
  {journal} {\bibinfo  {journal} {Physics of Plasmas}\ }\textbf {\bibinfo
  {volume} {12}},\ \bibinfo {pages} {063101} (\bibinfo {year}
  {2005})}\BibitemShut {NoStop}%
\bibitem [{\citenamefont {Biagioni}\ \emph {et~al.}(2016)\citenamefont
  {Biagioni}, \citenamefont {Anania}, \citenamefont {Bellaveglia},
  \citenamefont {Chiadroni}, \citenamefont {Cianchi}, \citenamefont
  {Di~Giovenale}, \citenamefont {Di~Pirro}, \citenamefont {Ferrario},
  \citenamefont {Filippi}, \citenamefont {Mostacci} \emph
  {et~al.}}]{biagioni2016electron}%
  \BibitemOpen
  \bibfield  {author} {\bibinfo {author} {\bibfnamefont {A.}~\bibnamefont
  {Biagioni}}, \bibinfo {author} {\bibfnamefont {M.}~\bibnamefont {Anania}},
  \bibinfo {author} {\bibfnamefont {M.}~\bibnamefont {Bellaveglia}}, \bibinfo
  {author} {\bibfnamefont {E.}~\bibnamefont {Chiadroni}}, \bibinfo {author}
  {\bibfnamefont {A.}~\bibnamefont {Cianchi}}, \bibinfo {author} {\bibfnamefont
  {D.}~\bibnamefont {Di~Giovenale}}, \bibinfo {author} {\bibfnamefont
  {G.}~\bibnamefont {Di~Pirro}}, \bibinfo {author} {\bibfnamefont
  {M.}~\bibnamefont {Ferrario}}, \bibinfo {author} {\bibfnamefont
  {F.}~\bibnamefont {Filippi}}, \bibinfo {author} {\bibfnamefont
  {A.}~\bibnamefont {Mostacci}},  \emph {et~al.},\ }\href@noop {} {\bibfield
  {journal} {\bibinfo  {journal} {Journal of Instrumentation}\ }\textbf
  {\bibinfo {volume} {11}},\ \bibinfo {pages} {C08003} (\bibinfo {year}
  {2016})}\BibitemShut {NoStop}%
\bibitem [{\citenamefont {Filippi}\ \emph {et~al.}(2018)\citenamefont
  {Filippi}, \citenamefont {Anania}, \citenamefont {Biagioni}, \citenamefont
  {Brentegani}, \citenamefont {Chiadroni}, \citenamefont {Cianchi},
  \citenamefont {Deng}, \citenamefont {Ferrario}, \citenamefont {Pompili},
  \citenamefont {Rosenzweig} \emph {et~al.}}]{filippi2018tapering}%
  \BibitemOpen
  \bibfield  {author} {\bibinfo {author} {\bibfnamefont {F.}~\bibnamefont
  {Filippi}}, \bibinfo {author} {\bibfnamefont {M.}~\bibnamefont {Anania}},
  \bibinfo {author} {\bibfnamefont {A.}~\bibnamefont {Biagioni}}, \bibinfo
  {author} {\bibfnamefont {E.}~\bibnamefont {Brentegani}}, \bibinfo {author}
  {\bibfnamefont {E.}~\bibnamefont {Chiadroni}}, \bibinfo {author}
  {\bibfnamefont {A.}~\bibnamefont {Cianchi}}, \bibinfo {author} {\bibfnamefont
  {A.}~\bibnamefont {Deng}}, \bibinfo {author} {\bibfnamefont {M.}~\bibnamefont
  {Ferrario}}, \bibinfo {author} {\bibfnamefont {R.}~\bibnamefont {Pompili}},
  \bibinfo {author} {\bibfnamefont {J.}~\bibnamefont {Rosenzweig}},  \emph
  {et~al.},\ }\href@noop {} {\bibfield  {journal} {\bibinfo  {journal} {Nuclear
  Instruments and Methods in Physics Research Section A: Accelerators,
  Spectrometers, Detectors and Associated Equipment}\ } (\bibinfo {year}
  {2018})}\BibitemShut {NoStop}%
\bibitem [{gpt()}]{gpt_web}%
  \BibitemOpen
  \href@noop {} {\enquote {\bibinfo {title} {{Pulsar Physics}},}\ }\bibinfo
  {howpublished} {\url{http://www.pulsar.nl/gpt}}\BibitemShut {NoStop}%
\end{thebibliography}%
\bibliographystyle{apsrev4-1}

\end{document}